\documentstyle[aps,prb,twocolumn,epsf,subfigure]{revtex}
\input{psfig}
\begin{document}
\draft
\def\sech{\hbox{sech}}
\def\mod{\hbox{ mod }}
\widetext
\title{Coherent Charge Transport in Metallic Proximity Structures}
\author{A.A.Golubov$^{a,b}$, F.K.Wilhelm$^c$ and A.D.Zaikin$^{c,d}$}
\address{$^a$ Institute of Thin Film and Ion
Technology, Research Centre J\"ulich (KFA), D-52425 J\"ulich, Germany\\ 
$^b$ Institute of Solid State Physics, 142432 Chernogolovka, Russia\\ $^c$
Institut f\"ur Theoretische Festk\"orperphysik, Universit\"at Karlsruhe,
76128 Karlsruhe, FRG \\ $^d$ I.E.Tamm Department of
Theoretical Physics, P.N.Lebedev Physics Institute, Leninskii prospect
53, 117924 Moscow, Russia }
\maketitle
\begin{abstract}
We develop a detailed microscopic analysis of electron transport in normal diffusive
conductors in the presence of proximity induced superconducting correlation.
We calculated the linear conductance of the system, the profile of the
electric field and the densities of states. In the case of transparent 
metallic boundaries the temperature dependent conductance 
has a non-monotoneous ``reentrant'' structure. We argue that this behavior
is due to nonequilibrium effects occuring in the normal metal in the
presence of both superconducting correlations and the electric field there.
Low transparent tunnel barriers suppress the nonequilibrium effects and
destroy the reentrant behavior of the conductance.  
If the wire contains a loop, the conductance shows Aharonov-Bohm 
oscillations with the period $\Phi_0=h/2e$ as a function
of the magnetic flux $\Phi$ inside the loop. The amplitude of these 
oscillations also demonstrates the reentrant behavior vanishing at $T=0$
and decaying as $1/T$ at relatively large temperatures. The latter
behavior is due to low energy correlated electrons which penetrate
deep into the normal metal and ``feel'' the effect of the magnetic flux $\Phi$.
We point out that the density of states and thus the ``strengh'' of the
proximity effect can be tuned by the value of the flux inside the loop. 
Our results are fully consistent with recent experimental findings.
\end{abstract}

\section{Introduction}

Recent progress in nanolithographic technology revived the interest to both
experimental and theoretical investigation of electron transport in various
mesoscopic proximity systems consisting of superconducting and normal
metallic layers. In such systems the Cooper pair wave function of a
superconductor penetrates into a normal metal at a distance which increases
with decreasing temperature \cite{dG}. At sufficiently low temperatures this
distance becomes large and the whole normal metal may acquire
superconducting properties. Although this phenomenon has been already
understood more than thirty years ago and intensively investigated during
past decades, recently novel physical features of metallic proximity systems
have been discovered \cite{Kas,Petr,Pot,Cour,Cour2,Cour3,Char} and studied theoretically (see 
\cite{Zai90,Volk92,Volk94,VZK,Been92,Been94,HN,Naz,Zaik,Spivak,Yip,Lambert,WSZ} and further
references therein).

In this paper we study the influence of the proximity effect on transport
properties of a diffusive conductor in the limit of relatively low
temperatures and voltages. We will assume that this conductor is brought in
a direct contact to a superconducting reservoir which serves as an effective
injector of Cooper pairs into a normal metal. We will show that
if the system contains no tunnel barriers there are two different physical
regimes which determine the system conductance in different temperature
intervals. It is well known that proximity induced superconducting correlation
between electrons in a diffusive normal metal survives at a distance of order $%
\xi_N \sim \sqrt{{\cal D}/T}$, where ${\cal D}=v_Fl_{imp}/3$ is the
diffusion coefficient. As $T$ is lowered the proximity induced superconductivity
expands into the normal metal and, consequently, the ``normally conducting''
part of the system effectively shrinks in size. This effect results in
increasing of the conductance of a normal metal. At sufficiently low
temperature the length $\xi_N$ becomes of order of the size of the normal
layer and the system behavior becomes sensitive to a physical choice of the 
boundary condition at the edge of the normal wire opposite to that attached 
to a superconductor. 

One possible choice of this boundary condition corresponds
to the assumption that a nontransparent 
barrier is present at the edge of this wire. Then electrons
cannot diffuse out of the wire, the proximity induced superconducting 
correlation survives everywhere in the system and a real gap in the
quasiparticle spectrum develops in the N-metal \cite{GK}. The value of this
gap is of order of $\varepsilon_g\sim \min (\Delta , {\cal D}/L^2)$, $\Delta$ is the
bulk superconducting gap and $L$ is the length of the normal wire. 

Another possible situation corresponds to the presence of a big normal 
reservoir N' directly
attached to the N-wire by means of a highly transparent contact. In this case 
even at very low $T$ the proximity induced Cooper pair amplitude is essentially
nonhomogeneous in the N-metal. Indeed, close to a superconductor this amplitude
is large, whereas in the vicinity of a normal reservoir it is essentially 
suppressed. Thus, strictly speaking, the whole N-wire cannot be characterized
by the real gap in its quasiparticle spectrum. In the absense of a 
potential barrier between N and N' this gap is 
obviously equal to zero at the NN' interface and -- as will be demonstrated
-- everywhere in the normal metal. Nevertheless, it turns out that 
the density of
states in the N-metal shows a soft pseudogap which is again of the order of 
$\varepsilon_g$. In other words, the spacially averaged normalized density of states $N_N(\varepsilon )$
in the N-wire at small $\varepsilon \lesssim \varepsilon_d$ is smaller than its normal 
state value $N_N<N(0)$ but always remains nonzero.
It increases with increasing $\varepsilon$ and reaches the 
value $N_N=N(0)$ at $\varepsilon \gtrsim \varepsilon_d$. This is the key
point for understanding the low temperature behavior of the conductance of our 
system.
As the temperature increases from zero, higher and higher values
of $\varepsilon$ contribute to the current and the system conductance
-- due to the increase of $N_N$ with $\varepsilon$ -- increases with $T$. 
This regime takes place until the temperature reaches the value $T \sim 
\varepsilon_d$ where the crossover to a high temperature behavior takes
place. Note that similar behavior of the normal metal conductance in the
presence of proximity-induced superconductivity has been recently found 
by Nazarov and Stoof \cite{NazSt}. 

An interesting feature of the system without tunnel barriers
is that at $T=0$ its conductance $exactly$ coincides with
that of a normal metal with no proximity effects. This result has been
first obtained by Artemenko, Volkov and Zaitsev \cite{AVZ} for the
case of a normal-superconducting constriction. Although it is 
already around for many years the physical meaning of this result -- if any --
still needs to be understood. At the first sight the linear conductance
of the system at $T=0$ should be smaller than in the normal state because
of the presence of the (pseudo)gap in the normal DOS $N_N$ 
at low energies. Why is this not the case?

In order to answer this question we should recall the well known fact
that in the presence of nonequilibrium effects the current flowing in a 
superconductor depends not only on the normal DOS but is 
characterized  by a set of generalized DOS \cite{Schmid}.
Our problem is just a particular example of a nonequilibrium superconductor:
on one hand superconducting correlation penetrates into the normal metal and
the Cooper pair amplitude is nonzero there, on the other hand in the
absence of low transparent tunnel barriers the electric field also penetrates
into the N-metal and drives the quasiparticle distribution function out
of equilibrium. We will argue that in this situation one of the generalized DOS
(below we define it as $N_S(\varepsilon )$) -- that is nonzero in the N-layer due to the presence of proximity induced superconducting 
correlation at low energies -- plays an important role and also contributes
to the system conductance. In other words, in the presence of the electric field
inside the system {\it both uncorrelated and correlated electrons contribute to a
dissipative current}. This is the reason why in the presence of proximity induced
superconductivity the system conductance is never smaller than its normal state
value although the normal DOS $N_N(\varepsilon )<N(0)$ at low energies \cite{FN}.

We would like to emphasize that the situation is entirely different in the 
presence of low transparent tunnel barriers. Provided their resistances are
much larger than that of the N-metal the whole voltage drop takes place at these
barriers and the electric field does not penetrate into the N-layer. In this case
only uncorrelated electrons contribute to the dissipative current and therefore
only the normal DOS $N_N$ matters. As a result the temperature dependence
of the system conductance changes. We will demonstrate that with lowering 
the barrier transparency the crossover
takes place to the effective conductance decreasing monotonously with T,
characteristic for two serial NIS' tunnel junctions (S' is now the diffusive
normal conductor with the proximity-induced gap). 

Note that both types of the
behavior, namely reentrant and monotonously decreasing with $T$
conductance have been observed in the experiments \cite{Petr,Cour3,Char}. Furthermore, we
would like to point out that both densities of states 
$N_N(\varepsilon )$ and $N_S(\varepsilon )$ can be investigated in one experiment.
We will come back to this point further below.

When the system contains a mesoscopic loop of a normal metal, the conductance as a
function of the magnetic flux through the loop shows oscillations with
the period $\Phi_0=h/2e$ (superconducting flux quantum). Although the Cooper 
pair amplitude (and thus the supercurrent) in the ring is exponentially
small at $T\gg\epsilon_d$, the amplitude of these oscillations 
decays only as $\propto 1/T$. This again illustrates an important difference
in the behavior of kinetic (conductance) and thermodynamic (supercurrent)
quantities. Below we will argue that in the systems considered here 
even at large $T \gg \epsilon_d$ the behavior of the first quantity is dominated 
by correlated low
energy electrons with $\epsilon \lesssim  \epsilon_d$ penetrating far into
the normal metal whereas the contribution of electrons with $\epsilon \sim T$
to the second one is only important. Again the presence of the electric field
inside the N-metal is crucially important for this effect.
 At low $T$ the oscillation amplitude again
shows the reentrant behavior and vanishes in the limit $T \rightarrow 0$ as $T^2$. 
A clear experimental evidence for a $1/T$ decay of the conductance oscillations
has been recently reported in Ref. \onlinecite{Cour2}.

Finally we point out that making use of the the geometry with a metallic
loop one can easily tune the densities of states of the system by applying
a magnetic flux $\Phi$ inside this loop. We will show that e.g. for $\Phi =\Phi_0/2$
the proximity effect in the normal region ``after'' the loop is completely
suppressed and the normal DOS $N_N(\epsilon )=N(0)$ there. This effect can be
investigated experimentally and used for further studies of proximity induced
superconductivity in normal metullic structures.

The structure of our paper is as follows. 
In Section II we briefly describe the general kinetic approach based on quasiclassical
Green functions in the Keldysh technique and define the physical quantities of interest.
Then a detailed analysis of this quantities (conductance -- Section III, DOS --
Section IV, elecric field -- Section V) will be presented. Sections VI and VII are
devoted to the extension of our analysis to the proximity systems containing 
mesoscopic normal metal loops with a magnetic flux. The main results of the 
present paper are summarized in Section VIII. Further details related to different
geometric realizations of the proximity systems with loops are presented in Appendix. 
\section{Kinetic analysis}

\subsection{General formalism}

\begin{figure}
    \centerline{\epsfxsize8cm \epsffile{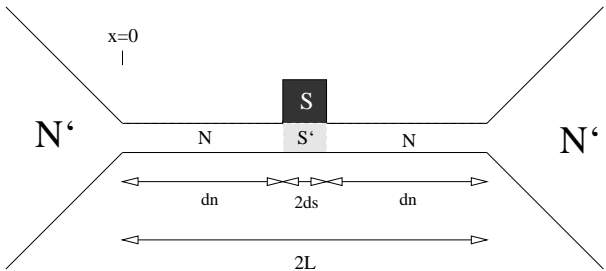}}
    \caption{The experimental system under consideration}
    \label{nns}
    \end{figure}

Let us consider a quasi-one-dimensional normal conductor of a length $2L$
with a superconducting strip of a thickness $2d_s$ attached to a normal
metal on the top of it and two normal reservoirs attached to its edges (see
fig.1). The length $L$ is assumed to be much larger than the elastic mean
free path $l_{imp}$ but much shorter than the inelastic one. This
geometrical realization has a direct relation to that investigated in the
experiments \cite{Petr,Cour,Char}. Two big normal reservoirs N' are assumed
to be in thermodynamic equilibrium at the potentials $V$ and $0$
respectively. In contrast to the case of a ballistic constriction \cite{BTK,ZGZ}
the potential drop within the system is distributed between the interfaces
and the conductor itself. The general approach to calculate the conductance 
of these structures was developed in \cite{Zai90,Volk92,VZK}. In what
follows we shall apply this method to analyse the temperature dependence of
the NS proximity structure of fig. \ref{nns}.

Such an experimental realization allows to prepare a
structure without effective tunnel barriers in the direction 
of the current flow. Even with ``perfect'' samples in a usual 
sandwich geometry, a
natural barrier shows up due to the inevitable mismatch of Fermi velocities
between different materials. This could well be one of the reasons why
in the previous experiments with sandwich-like structures, 
the reentrant behavior of the conductance was not detected \cite{Magnee}.

The electron transport through the metallic system can be described by the
equations for a matrix of quasiclassical Green functions $\stackrel{\vee }{G%
\text{ }}$in the contact \cite{Eliash,LO}: 
\begin{equation}
\label{G}\stackrel{\vee }{G\text{ }}=\left( 
\begin{array}{cc}
\stackrel{\wedge }{G}^R & \stackrel{\wedge }{G}^K \\ 0 & \stackrel{\wedge }{G%
}^A 
\end{array}
\right) 
\end{equation}
where $\stackrel{\wedge }{G^A}$, $\stackrel{\wedge }{G^R}$ and $\stackrel{%
\wedge }{G^K}$ are respectively the impurity-averaged advanced, retarded and
Keldysh Green functions. These functions are in turn matrices in the Nambu
space: 
$$
\stackrel{\wedge }{G^R}=\stackrel{\wedge }{\sigma _z}g^R+\stackrel{\wedge }{%
i\sigma _y}f^R,\stackrel{\wedge }{G^A}=-(\stackrel{\wedge }{G^R})^{*}%
\hbox{and}\stackrel{\wedge }{G^K}=\stackrel{\wedge }{G^R}\stackrel{\wedge }{f%
}-\stackrel{\wedge }{f}\stackrel{\wedge }{G^A}. 
$$
Here the distribution function $\stackrel{\wedge }{f}=f_l+\stackrel{\wedge 
}{\sigma _z}f_t$, where $f_l=\tanh (\varepsilon /2T)$ and $f_t$describes
deviation from equilibrium. Taking advantage of the normalization
condition for the normal and the anomalous Green functions $%
(g^R)^2-(f^R)^2=1 $ it is convenient to parametrize $g^R=\cosh \theta ,$ $%
f^R=\sinh \theta $, where $\theta \equiv \theta _1+i\theta _2$ is a complex
function. Deep in the bulk superconductor it is equal to $\theta _s=1/2\ln
\left[ (\Delta +\varepsilon )/(\Delta -\varepsilon )\right] -i\pi /2$ for $%
\varepsilon <\Delta $ and $\theta _s=(1/2)\ln \left[ (\varepsilon +\Delta
)/(\varepsilon -\Delta )\right] /2$ for $\varepsilon >\Delta $ (here and
below we omit the indices R(A)).

The current $I$ and the electrostatic potential $\phi$ are
expressed through $\stackrel{\vee }{G\text{ }}$ as 
\begin{equation}
\label{curr-d}I=\frac{\nu {\cal D}S}2\int_{-\infty }^\infty d\varepsilon 
\text{ }Sp\left[ \stackrel{\wedge }{\sigma _z}\stackrel{\vee }{G\text{ }}%
\partial _x\stackrel{\vee }{G\text{ }}\right] ^K, 
\end{equation}
\begin{equation}
\phi(x)=\int_0^\infty d\varepsilon\,\hbox{Tr}\hat{g}^K(x,\varepsilon)
=\int_0^\infty d\varepsilon\,f_t(x,\varepsilon)\nu_\varepsilon (x), 
\end{equation}
where $\nu $ is the the density of states, $\nu_\varepsilon(x)=\Re(g^R_%
\varepsilon(x))$ and $S$ is the crossection area of $N$ conductor.

Being expressed in terms of the function $\theta (\varepsilon ,x)$ the
equations \cite{Eliash,LO} for the Green functions and the distribution
function for the N-metal take a particularly simple form 
\begin{equation}
\label{eq1}{\cal D}\partial _x^2\theta +2i\varepsilon \sinh \theta =0 
\end{equation}

\begin{equation}
\label{eq2}\partial _x\left[ {\cal D}(\cosh ^2\theta _1)\partial
_xf_t\right] =0, 
\end{equation}
$x$ is the coordinate along the N-conductor. Here we neglected the processes
of inelastic relaxation and put the pair potential in the normal metal equal
to zero $\Delta _N=0$ assuming the absence of electron-electron interaction
there.

Before we come to a detailed solution of the problem let us point out that
the conclusion about the anomalous behavior of the system conductance can be
reached already from the form of eq. (\ref{eq2}). Indeed it is quite clear
from (\ref{eq2}) that the effective diffusion coefficient
${\cal D}_{\hbox{eff}}={\cal D}\cosh^2\theta_1$ increases in the
N-regions with proximity-induced superconductivity and, therefore, the
electric field is partially expelled from these regions. This energy
dependent field modulation is controlled by the solution for $\theta
(\varepsilon ,x)$ and is directly related to the physical origin of the
anomalous temperature dependence of the system conductance discussed below.

The equations (\ref{eq1}) and (\ref{eq2}) should be supplemented by the
boundary conditions at the interfaces of the normal metal N. Assuming that
the anomalous Green function of big normal reservoirs N' is equal to zero
from \cite{KL,Volk92} we obtain 
\begin{equation}
\label{BC}
\begin{array}{c}
\xi _N^{*}\gamma _B\partial _x\theta =\pm \sinh \theta , \\ 
\xi _N^{*}\gamma _B\cosh \theta _1\partial _xf_t=\pm \cosh \theta _2(f_t-f_t(x=0,2L)),
\end{array}
\end{equation}
where $\gamma _B=R_b/\rho _N\xi _N^{*}$ is the interface resistance
parameter, $R_b$ is the specific resistance of the interface between the
N-conductor and the N'-reservoirs, $\rho _N$ is the resistivity of the
N-metal, and $\xi _N^{*}=\sqrt{D_N/2\pi T_c}$ is the temperature independent
characteristic length scale in N (note that the coherence length in N $\xi
_N(T)=\sqrt{D_N/2\pi T}$ is T-dependent). 

In general we should also fix the boundary condition at the interface
between the N-metal and the superconductor. For the case of a perfect
transparency of this interface and for
typical thickness of the normal layer $w_N\sim \sqrt{S}$, Cooper
pairs easily penetrate into it due to the proximity effect and the Green
functions of the N-metal at relatively low energies for $d\le x\le d+2d_s$
are equal to those of a bulk superconductor $\theta =\theta _s$ (the
influence of finite transparency of the NS-contact will be discussed below).
In this sence the region of a normal metal situated directly
under the superconductor can be also treated as a piece
of a superconductor S' and the solution of (\ref{eq1}), (\ref{eq2}) needs to
be found only for $0<x<d$ (without loss of generality we will stick to a
symmetric configuration). 

Cooper pairs penetrate into the normal conductor
also in the case of a not perfectly transparent NS interface. 
As it is demonstrated below, the energy gap is induced in S' region in this case.
As a result, for a sufficiently long N-wire, which is only considered here,
the presence of the barrier at the NS interface will not influence the results 
derived for the system conductance.

\subsection{Physical quantities of interest}

Proceeding along the same lines as it has been done in ref. \cite{VZK} we
arrive at the final expression for the current 
\begin{equation}
\label{current}I=\frac 1{2R}\int_0^\infty d\varepsilon \left[ \tanh
\left({\varepsilon +eV\over2T}\right)-\tanh\left({\varepsilon -eV\over2T}\right)\right] D(\varepsilon ), 
\end{equation}
where $D(\varepsilon )$ defines the effective transparency of the system 
\cite{VZK} 

\begin{equation}
D(\varepsilon )=\frac{1+r}{\frac r{\cosh \theta _1(x=0,\varepsilon )\cos
\theta _2(x=0,\varepsilon )}+\frac 1L\int_0^Ldx\,\sech^2\theta
_1(x,\varepsilon )},
\label{D(E)} 
\end{equation}
$R=R_b+R_N$ and $r=R_b/R_N\equiv \gamma _B\xi _N^{*}/L$ , $R_N$ is
the resistance of the N-metal.

Let us consider the case of a sufficiently long normal conductor $d^2\gg {\cal %
D}/\Delta $. Then at low temperatures $T\ll \Delta $ the interesting energy
interval is restricted to $\varepsilon \ll \Delta $. For such values of $%
\varepsilon $ the contribution of the $S^{\prime }$-part of the normal
conductor shows no structure and can be easily taken into account with the
aid of obvious relations 
\begin{equation}
\int_0^Ldx\,\sech^2\theta _1(x,\varepsilon )=\int_0^ddx\;\sech^2\theta
_1(x,\varepsilon )+d_s\sech^2\theta _{s,1}
\end{equation}
and $\sech^2\theta _{s,1}=\left( 1-{\frac{\varepsilon ^2}{\Delta ^2}}\right) 
$ (no barrier at the NS interface) or $\sech^2\theta _{s,1}=(1-\frac{%
\varepsilon ^2}{\Delta _{gN}^2})$ (the barrier is present at the NS
interface). Due to this reason we will discuss only the properties of the $N$%
-part ($0<x<d$). For the sake of completeness we will also demonstrate the
effect of finite $d_s$ in the end of our calculation.

For the differential conductance of the $N$-part $0\le x\le d$ normalized to
its normal (``non-proximity'') value in the zero bias limit eq. (\ref
{current}) yields 
\begin{equation}
\label{conduct}\bar{G}_N=\left({\frac{RdI}{dV}}\right)_{V=0}={\frac{1}{%
2T}}\int_0^\infty d\varepsilon{D(\varepsilon)\sech^2(\varepsilon/2T)}. 
\end{equation}
Analogously the normalized zero-bias electrostatic potential distribution
reads 
\begin{eqnarray}
\nonumber\phi_0(x)=\lim_{V\rightarrow0}{\frac{\phi(x)}{V}}&=&{\frac{1%
}{2Td}}\int_0^\infty d\varepsilon\, D(\varepsilon)\nu_\varepsilon(x)\sech%
^2(\varepsilon/2T)\times\\
&&\times\int_x^{d}dx^\prime\,\sech^2(\theta_1(x^\prime)) \label{potential}
\end{eqnarray}

The normal density of states is given by the normal Green's
function via the standard relation $\nu_\epsilon(x)=N(0)\Re (g_\epsilon(x))$
which enters into the conductance in the form $\cosh^2\theta_1=(\Re
g)^2+(\Im f)^2$ together with a 'correlation DOS'
$\eta_\epsilon(x)=N(0)\Im (f_\epsilon(x))$. The importance of the
latter quantity for understanding the effects discussed here has been 
already pointed out in the Introduction. We will discuss the features
of these local densities as well as the averaged ones:

\begin{eqnarray*}
N_N(\epsilon)&=&\int d\bar{x}\nu_\epsilon(\bar{x})\\
N_S(\epsilon)&=&\int d\bar{x}\eta_\epsilon(\bar{x})\\
\end{eqnarray*}

As it has been already mentioned the ``correlation DOS'' $\eta$ belongs to 
the set of generalized densities of
states familiar from the standard theory of nonequilibrium superconductivity
\cite{Schmid,Beyer}. It reflects the presence
of superconducting correlations at low energies.
E.g. in a BCS superconductor this function reads 
$\eta={\Delta\Theta(\Delta-\epsilon)\over\sqrt{\Delta^2-\epsilon^2}}$.
In our case this function is not only energy- but also space-dependent
due to the fact that the proximity induced superconducting correlation
decays inside the normal metal. But the physical meaning of it remains 
the same as in standard nonequilibrium superconductivity 
theory \cite{Schmid}: $\eta$ plays a role 
whenever the quasiparticle distribution function of a superconductor is
driven out of equilibrium. It happens e.g. in the well-known problems of 
charge relaxation \cite{Beyer} and imbalance \cite{Schoen}. It happens
also here due to a simultaneous presence of the electric field and the
proximity induced superconducting correlation in the normal metal.

\subsection{Influence of finite barrier transparency at the top NS-interface}

Let us consider the effect of a tunnel barrier at the NS interface 
in more details.

Under the assumption that the N-wire thickness is small $w_N\ll \xi _N^{*}$ the equation for $\theta $ in the region 
$0\leq x\leq 2d_s$ underneath the superconducting terminal can be derived by
the method of Ref. \cite{Kupr-89}: 
\begin{equation}
\label{teta-B}{\cal D}\partial _x^2\theta +2i\widetilde{\varepsilon }\sinh
\theta +\widetilde{\Delta }\cos \theta =0
\end{equation}
where the effective order order parameter $\widetilde{\Delta }=\sin \theta
_s/\gamma _B^{NS}$, the effective energy $\widetilde{\varepsilon }%
=\varepsilon +\cos \theta _s/\gamma _B^{NS}$, $\gamma _B^{NS}=(R_b/\rho
_N)(w_N/\xi _N^{*2})$ is the interface transparency parameter. Here $\theta
_s$ is the solution in S which is set equal to the bulk value $\tan
{}^{-1}(i\Delta /\varepsilon )$, a good approximation for thin N film $%
w_N\ll \xi _N^{*}$. With these substitutions the equation (\ref{teta-B}) in
the N film has the form similar to that in a superconductor. This equation
is valid for $\gamma _B^{NS}>(w_N/\xi _N^{*})^2$ i.e. for sufficiently small
transparency of the NS interface: $\left\langle D\right\rangle <l_N/w_N$. 

As follows from Eq.(\ref{teta-B}), superconducting properties of the N-layer
are described in terms of the energy-dependent coherence length
\begin{equation}
\label{xi-eps}\xi _N(\epsilon )=\left\{ \hbar D_N/2\left[ \gamma
_B^{-2}-\epsilon ^2-2i\gamma _B^{-1}\epsilon \cos \theta _S\right]
^{1/2}\right\} ^{1/2},
\end{equation}
which determines an exponential decay of $N_N(x ,\epsilon )$ with $x $ .

The expression for $\xi_N(\epsilon )$ has a pole 
at the gap energy $\epsilon =\Delta _{gN}$, which
signals the decay of quasiparticles entering N at $\epsilon <\Delta _{gN}$.
At high energy $\epsilon \gg \pi T_c/\gamma _B^{NS}$ the well known result $%
\xi _N(\epsilon )=(\hbar D_N/2\epsilon )^{1/2}$ is reproduced, whereas at
low energies $\epsilon \ll \pi T_c/\gamma _B^{NS}$ one obtains $\xi
_N=(\hbar D_N\gamma _B^{NS}/2\pi T_c)^{1/2}$. Thus the effective length
scale in N increases with the decrease of the SN interface transparency.

It is straightforward to calculate the gap energy $\Delta _{gN}$ assuming the
``rigid'' boundary conditions $\theta _s=\tan {}^{-1}(i\Delta /\varepsilon )$
either from the pole of $\xi _N(\epsilon )$ or, making use of the solution $\theta
=\tanh ^{-1}\left[ \sinh \theta _s/(\cosh \theta _s-i\varepsilon \gamma
_B^{NS})\right] $ and calculating the quasiparticle density of states
$N_N(\varepsilon )=Re\cosh \theta $.
Subsituting the expression for $\theta _s$ into this solution 
one arrives the equation for the energy gap

\begin{equation}
\label{gap}t^3+2Ct^2+(C^2-1)t-2C=0,
\end{equation}
where $t=\sqrt{1-(\Delta _{gN}/\Delta )^2}$ and $C=\pi T_c/\gamma
_B^{NS}\Delta $. The general solution is rather cumbersome, therefore here 
we present only its asymptotic forms. 

\begin{figure}
   \vspace{-20mm}
    \centerline{\epsfxsize12cm \epsffile{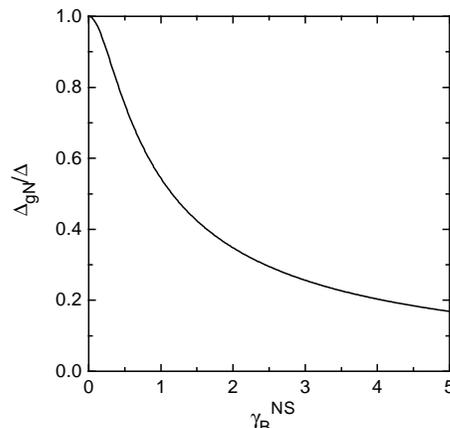}}
    \vspace{-80mm}
    \caption{Reduced gap on the normal side of the $NS$-boundary in
   the presence of barriers}
    \label{deltan}
    \end{figure}

The gap is given by $\Delta
_{gN}/\Delta =1-2(\gamma _B^{NS}\Delta /\pi T_c)^2$ for large transparency
of the NS interface, $\gamma _B^{NS}\Delta /\pi T_c\ll 1$, and by $\Delta
_{gN}=\Delta /(1+\gamma _B^{NS}\Delta /\pi T_c)$ for small transparency, $%
\gamma _B^{NS}\Delta /\pi T_c\gg 1$.  In the latter case (the McMillan
limit), the expression for the gap may be written as $\Delta _{gN}=\pi
T_c/\gamma _B^{NS}\equiv \hbar v_{FN}\left\langle D\right\rangle /4w_N$. The
gap originates from the finite average lifetime $\tau _N=2w_N/\left\langle
D\right\rangle v_{FN}$ for quasiparticles in the N layer with
respect to Andreev scattering from S, since a contribution of gapless quasiparticle
trajectories parallel to the NS interface is eliminated in the diffusive
regime. The dependence of $\Delta _{gN}$ on $\gamma _B^{NS}$ in the whole
range of transparencies is presented in Fig.\ref{deltan}. Note that under a 
substitution
$w_N=\pi /k_F$ the above expressions reproduce the result of Ref.\cite{Volk-95} for
a gap induced in a 2D electron gas in contact to a superconductor.
We also note that in the
case of two superconducting terminals attached to N
the gap aquires a phase factor $\Delta _{gN}\rightarrow \cos
(\varphi /2)\Delta _{gN}$, $\varphi $ being the phase difference between
the terminals.

At the subgap energies $\varepsilon \leq \Delta _{gN}$ the solution in N is $%
\theta \simeq $ $1/2\ln \left[ (\Delta _{gN}+\varepsilon )/(\Delta
_{gN}-\varepsilon )\right] -i\pi /2$ whereas above the gap, at $\varepsilon
>\Delta _{gN}$, $\theta \simeq 1/2\ln \left[ (\varepsilon +\Delta
_{gN})/(\varepsilon -\Delta _{gN})\right] $, in complete analogy with the
solutions in a bulk superconductor but with reduced gap. Therefore the
results for conductance of N depend on the relation between the gap $\Delta _g$
and the Thouless energy $\varepsilon _d=D/d^2$. Since for the case
of a sufficiently long N-wire, which is only considered in this paper, the
condition $\Delta _{gN}\gg $ $\varepsilon _d$ is satisfied, the
presence of the barrier at the NS interface does not influence our results
for the system conductance.

\section{Conductance}

\subsection{Perfectly transparent boundaries}

The analysis of the problem can be significantly simplified in the case of
perfectly transparent interfaces ($\gamma _B=0$). In this case the boundary
conditions are 
\begin{eqnarray}
\label{bc1}
\theta(0)&=&0\\
\theta(d)&=&\theta_S
\end{eqnarray}
for the contact to the normal and the superconducting reservoir
respectively. The effective transparency of the N-part then reads 
\begin{equation}
D(\varepsilon )=\left( {\frac 1d}\int_0^ddx\;\sech^2(\theta _1(x))\right)
^{-1}.
\end{equation}

As it was already pointed out for relatively long normal conductors and at
low $T$ only the energies $\varepsilon \ll \Delta $ give an important
contribution to the conductance. In this case the typical energy scale is
defined by the Thouless energy $\epsilon _d={{\cal D}/d^2}\ll \Delta ,\Delta
_{gN}$. For these energies we can set $\theta _S=-i{\pi /2}$. Let us first
put $T=0$. Then the thermal distribution factor $\sech^2(\varepsilon /2T)/(2T)$ reduces to a delta function and we have 
\begin{equation}
\bar G_N(T=0)=D(0),
\end{equation}
i.e. we only need the solution of (\ref{eq1}) with boundary conditions (\ref
{bc1}) at $\varepsilon =0$, which is $\theta =-i{\frac \pi 2}{\bar
x}$. This does not depend on ${\cal D}$, so the correlations are
destroyed by the influence of the boundary conditions but not by
thermal excitation or by impurity scattering. From here, we can
calculate the conductance
\begin{equation}
\bar{G}_N(T=0)=1,
\end{equation}
i.e. at $T=0$ the system conductance exactly coincides with its normal state
value (cf. \cite{AVZ,NazSt}. This result, however, by no means implies the
destruction of the proximity induced superconductivity in the N-layer. 
Later on, we will demonstrate, that the DOS 
and the electrical field are completely different from their
values in the normal state and in fact only due the additional contribution
of correlated electrons the system conductance does not fall below its
normal state value. 


In the case $T\ll
\varepsilon _d$ we can calculate $\theta $ perturbatively. From ${{\cal D}%
\partial _x^2\theta =-2i\varepsilon \sinh \theta _0(x)}$ and (\ref{bc1}) we
get 
$$
\theta =-{\frac 8{\pi ^2}}{\frac \varepsilon {\varepsilon _d}}[\bar x-\sin (%
\bar x\pi /2)]-i{\frac \pi 2}\bar x. 
$$
Keeping only leading order terms in $\frac \epsilon {\epsilon _d}$, we get 
\begin{equation}
\bar \sigma _N=1+A{\frac{T^2}{\epsilon _d^2}},
\end{equation}
where $A={\frac{64}{3\pi ^4}}\left( {\frac 56}-{\frac 8{\pi ^2}}\right)
\approx 0.049$ is a universal constant. This means, that for low
temperatures $\bar G_N(T)$ grows quadratically on the scale of $%
\varepsilon _d$ and approaches the crossover towards the high temperature
regime discussed below.

In the limit $T\gg\epsilon_d$ (where we still have $T\ll\Delta$), the
contribution of the low energy components to the thermally weighted
integral for $\bar{G}_N(T)$ is $\propto 1/T$ as we will see below and
can therefore be neglected. We only
have to take into account the solutions of (\ref{eq1}) for energies $%
\epsilon\gg\epsilon_d$. It is well known (see e.g. \cite{VZK}), that for
this energy range the solution of (\ref{eq1}) together with (\ref{bc1})
reads 
\begin{equation}
\label{HochT}
\tanh(\theta(\bar{x})/4)=\tanh\left({\frac{i\pi}{8}}\right)e^{k(\bar{x}-1)} 
\end{equation}
where $k=d\sqrt{-2i\varepsilon/{\cal D}}$. By using obvious substitutions
and multiple-argument relations for hyperbolic functions, we arrive at the
following identity: 
\begin{equation}
\label{help}{\int_{\bar{x}}^1}d\bar{x}\,\sech^2(\theta_1(\bar{x}))=(1-\bar{x}%
)-4\sqrt{\frac{\varepsilon_d}{\varepsilon}}\int\limits_0^{\Re(k)(1-\bar{x})}
{q(y)\, dy\over(1+q(y))^2} 
\end{equation}
where $q(y)=4(3+2\sqrt{2}){e^{-2y}\sin^2y/\left(e^{-2y}+3+2\sqrt{2}\right)^2}
$.

For calculating $D(\varepsilon)$ we can, as the integrand becomes
exponentially small for $y\ge\Re(k)\gg1$, take the upper bound to infinity,
such that it becomes a universal constant. From there we can calculate the
conductance in this limit 
\begin{equation}
\label{root}\bar{G}_N(T)=1+B\sqrt{\frac{\varepsilon_d}{T}}, 
\end{equation}
where again $B=0.42$ is a universal constant. 

These results has a simple physical interpretation. Superconductivity
penetrates into the normal part up to $\xi_N=\sqrt{\frac{D}{2\pi T}}$,
whereas the rest stays normal, so the total voltage drops over a reduced
distance $d-\xi_N$. Thus the resistance of the structure is reduced
according to the Ohm law. In terms of the conductance, this means 
\begin{equation}
\bar{G}_N=1+B^\prime{\frac{\xi_N}{d}} 
\end{equation}
which is equivalent to (\ref{root}).

   \begin{figure}
    \centerline{\psfig{figure=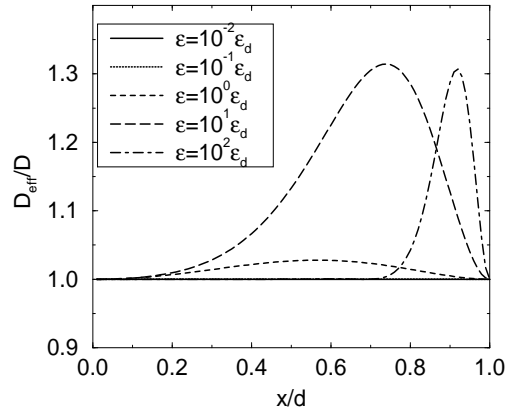,width=7cm,height=6cm}}
    \caption{Local effective diffusion constant}
    \label{emm}
    \end{figure}

Let us point out, that at both edges of the N-metal the local effective
diffusion constant ${\cal D}_{\hbox{eff}}=\cosh^2\theta_1{\cal D}$ is not 
enhanced (see Fig. \ref{emm}) in comparison to its normal state value, because 
either the Cooper pair amplitude
(at the NN' boundary) or the electric field (at the NS boundary) is equal
to zero due to the imposed boundary conditions. Inside the N-metal the value 
${\cal D}_{\hbox{eff}}$ becomes higher due to nonequilibrium effects in the
presence of superconducting correlations ($\eta \neq 0$). This effect is
small at very low energies and becomes more pronounced at $\varepsilon 
\sim \varepsilon_d$.

For temperatures comparable to $\varepsilon_d$ the problem was treated
numerically. The results show an excellent agreement with our analytical
expressions obtained in the corresponding limits.

    \begin{figure}
     \centerline{\psfig{figure=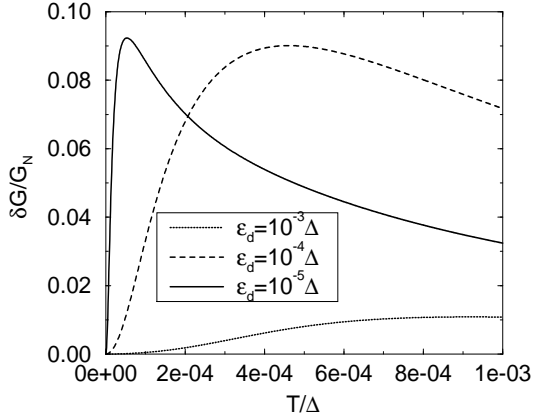,width=7cm,height=6cm}}
    \caption{Conductance in the case of transparent barriers}
    \label{conduc}
    \end{figure}

\begin{figure} 
    \centerline{\psfig{figure=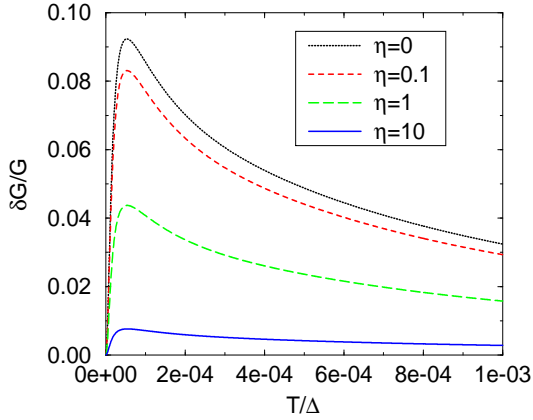,width=7cm,height=6cm}}
    \caption{Conductance normalized to the total length of the normal
    wire. $\eta:=d_S/d$, $\epsilon_d=10^{-5}\Delta$}
    \label{eta}
    \end{figure}

The numerical results (see fig. \ref{conduc}) confirm, that for
$\varepsilon_d\ll\Delta$ 
the universal scaling with $\frac{T}{\varepsilon_d}$ is excellently
fulfilled, the conductance peak with the height of about 9\% takes place at $%
T\approx5\varepsilon_d$ (cf. \cite
{NazSt}). This peak becomes smaller if we take into account the influence of
finite $d_S$ keeping $d$ fixed (fig. \ref{eta}) The qualitative features, however,
remain the same.

\subsection{Tunnel barriers}

\begin{figure}
    \centerline{\epsfxsize12cm \epsffile{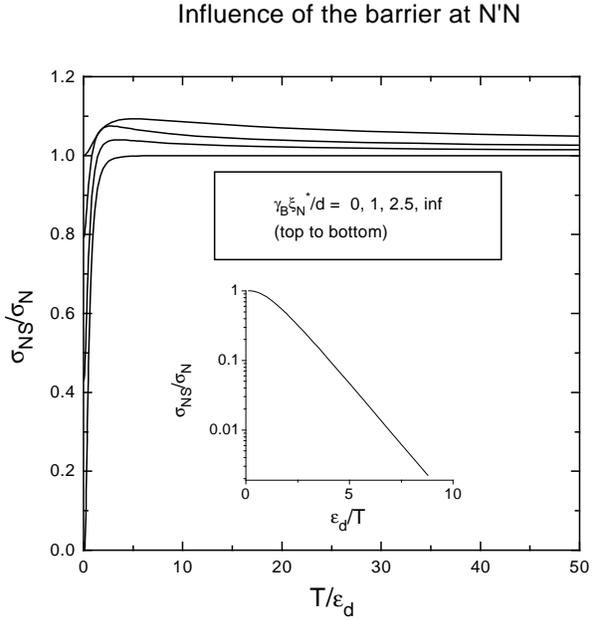}}
    \vspace{-60mm}
    \caption{Conductance in the presence of tunneling barriers}
    \label{contunn}
    \end{figure}

Let us now assume that a tunnel barrier is present at the N'-N interface. 
If one lowers the transparency of this barrier the crossover takes place to the
behavior demonstrating monotonously decreasing conductance with T
(Fig. \ref{contunn}),
which is typical for two serial NIS tunnel junctions. Fig. \ref{contunn} 
demonstrates the crossover with increasing $r=\gamma
_B\xi _N^{*}/d$. Inset shows the Arrenius plot for the case of $\gamma
_B\xi _N^{*}/d\gg 1$ which illustrates the activated tunnel-like behavior.

Formally this is due to the
term $r/\cosh \theta _1(x=0,\varepsilon )\cos \theta _2(x=0,\varepsilon )$
in the denominator of Eq.(\ref{D(E)}), i.e. the contribution of the barrier at the
N'-N interface. In the small transparency limit $r\gg 1$ the expression (%
\ref{D(E)}) reduces to the standard tunnel formula. The physical reason for this
behavior is transparent. For $r \ll 1$ the presence of a tunnel barrier is not
important, the electric field penetrates inside the normal metal and we come
back to the picture discussed above for perfectly transparent boundaries in which
case both normal and correlation DOS play a significant role. If,
however, the resistance of a tunnel barrier dominates over the Drude resistance
of the normal metal $r \gg 1$, the whole voltage drop is concentrated at the 
barrier, nonequilibrium effects in the N-metal are absent and therefore only
the normal density of states enters into the system conductance. 

An additional effect
is that a real gap instead of a soft pseudogap develops in the case low 
transparent tunnel barriers. The crossover between these two regimes is
discussed in more detail below.

Note that both types of behavior, namely nonmonotoneous and monotonously
decreasing with T conductance have been observed in the experiments \cite
{Petr}. 

    \begin{figure}
    \vspace{-10mm}
    \centerline{\epsfxsize12cm \epsffile{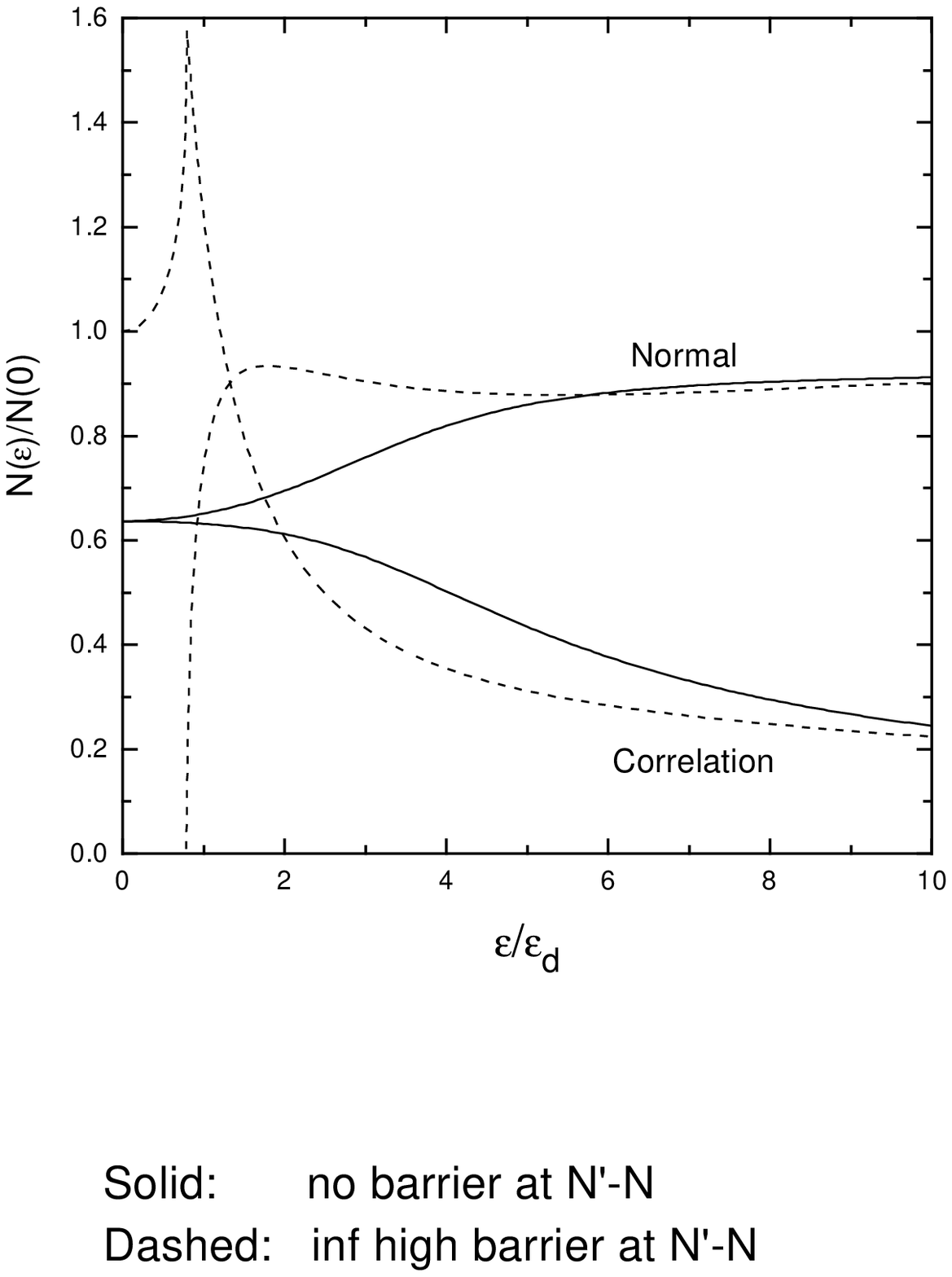}}
    \vspace{-45mm}
    \caption{Averaged normal and correlation DOS}
    \label{dos1}
    \end{figure}

\section{Density of states}

\subsection{Averaged density}

From our approximative solutions of the preceeding sections, the
densities of states can be easily calculated. For $\epsilon=0$ we have
$N_N=N_S=N(0){2\over\pi}$. At low energies $\epsilon\ll\epsilon_d$
there are quadratic corrections: $N_{N/S}=N(0)\left({2\over\pi}\pm
A_{1/2}\left({\epsilon\over\epsilon_d}\right)^2\right)$ with
$A_1={64\over\pi^5}\left(1-{8\over\pi^2}\right)\approx0.0396$ and
$A_2={16\over\pi^4}\left(1+{2\over\pi^5}\right)\approx0.198$. For high
energies, the densities approach their normal values, again with
square-root corrections
$N_N=N(0)\left(1-B_1\sqrt{\epsilon_d\over\epsilon}\right)$ and
$N_S=B_2\sqrt{\epsilon_d\over\epsilon}$ with $B_1\approx 0.321$ and
$B_2\approx0.75$.

Together with our numerical data (see Fig. \ref{dos1}), this demonstrates
the presence of a soft pseudogap in the density of states below the
energy $\epsilon_d$. Similar results have also been discussed in \cite{GK}. 

    \begin{figure}
    \vspace{-15mm}
    \centerline{\epsfxsize8cm\epsffile{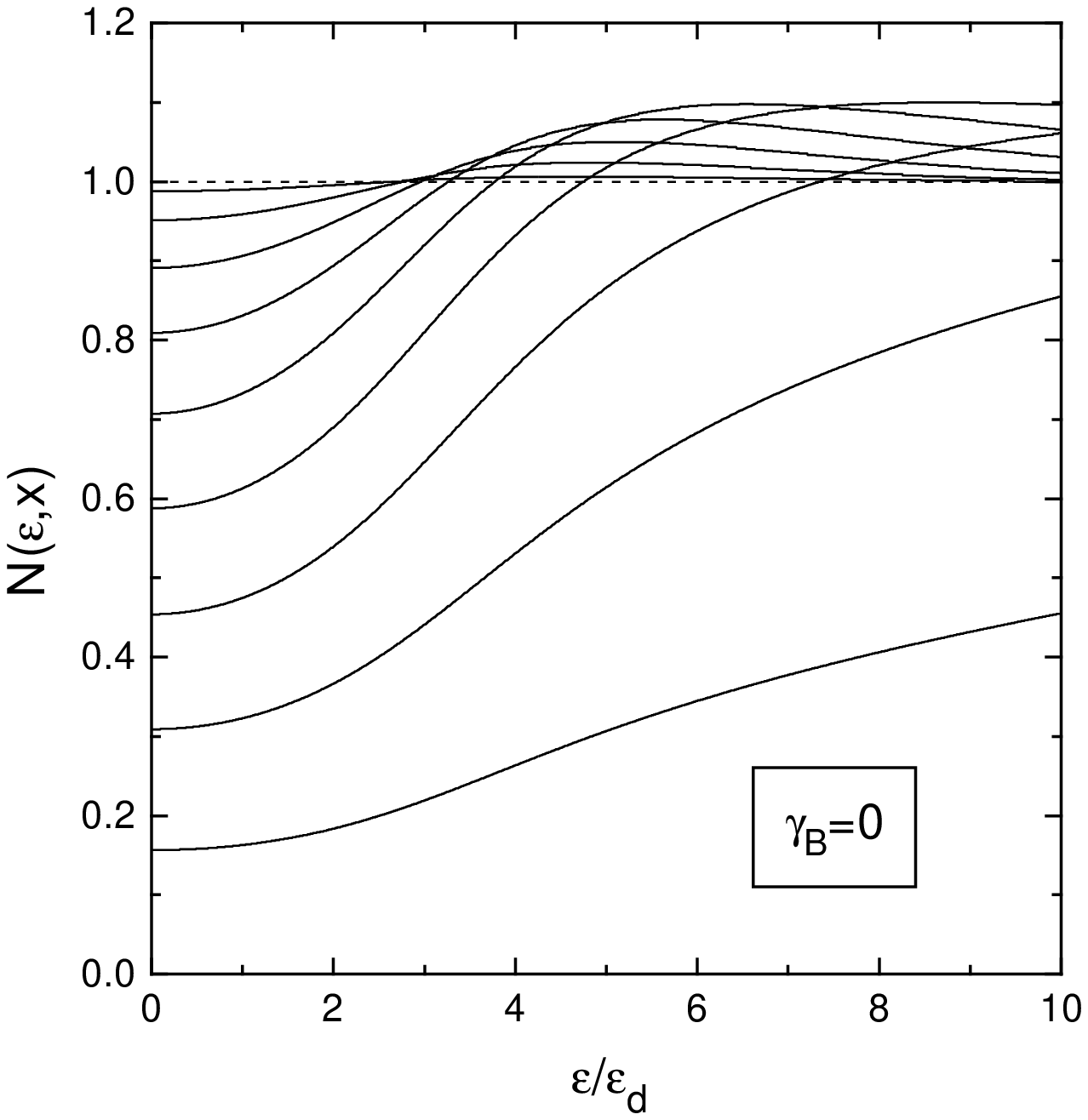}}
    \vspace{-55mm}
    \centerline{\epsfxsize8cm \epsffile{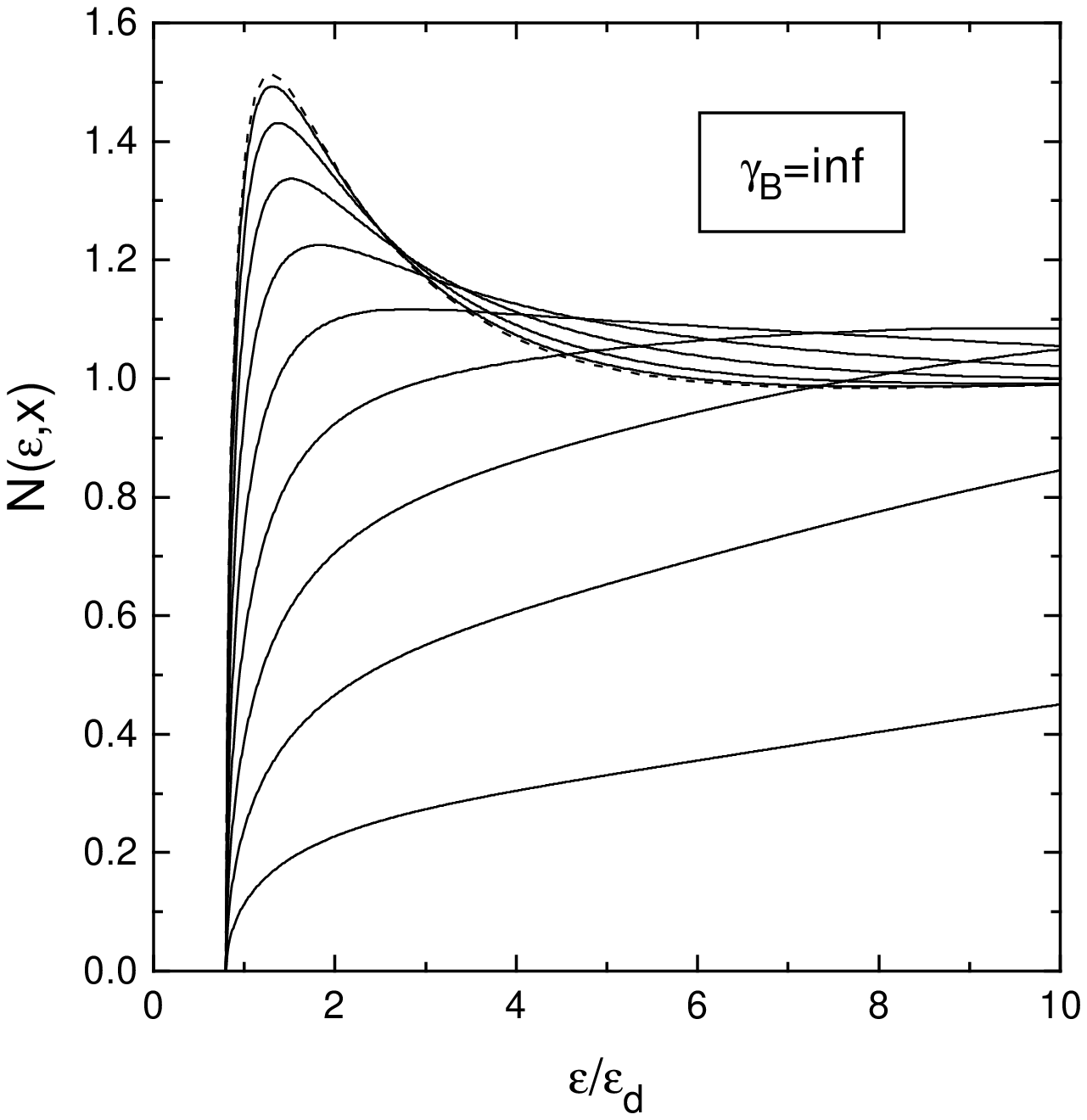}}
    \vspace{-35mm}
    \caption{Local DOS for different N-N$^\prime$-boundaries. Top:
    Transparent, Bottom: Non-transparent}
    \label{lodo}
    \end{figure}

\subsection{Spatial dependence and tunneling experiments}

It is also interesting to investigate the spatial dependence of the DOS in 
the normal layer. 
Fig. \ref{lodo} show local normal DOS $N_N$
calculated for perfectly transparent ($\gamma _B=0$) and nontransparent 
($\gamma _B=\infty $) NN' interfaces, respectively, at different
distances from the NS boundary: $x/d$= 0.1, 0.2, ..., 1. The difference
between these two cases is quite obvious: whereas for $\gamma _B=0$ the normal
DOS at low energies is always finite, becoming larger at larger values of $x$,
for $\gamma _B=\infty $ a real gap in the density of states clearly shows up
at all energies. Similar results have been recently discussed in Refs. 
\cite{aminov,golubov,belzig}. The overall behavior of the local correlation 
DOS at each value of $x$ is similar to its average value.  

\begin{figure}
     \centerline{\psfig{figure=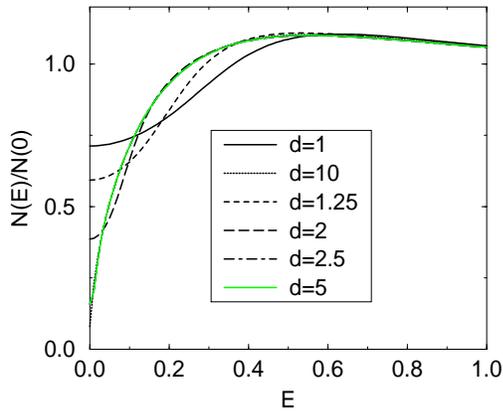,width=7cm,height=6cm}}
    \caption{Size effect on the local normal DOS. Here, the density of
    states at a fixed distance $x=0.5$ from the $NS$-boundary is
    plotted for different values of the total length $L$ of the N-part.}
    \label{sizeeff}
    \end{figure}

It is important to emphasize that both $\nu (x)$ and $\eta (x)$ are measurable quantities and 
can be directly probed in experiments. Recently the spatial and energy
dependence of the normal DOS has been studied in tunneling experiments 
\cite{gueron}. The data \cite{gueron} show a qualitative agreement
with theoretical predictions. The results obtained here suggest that 
much better agreement can be achieved if one takes into account 
smearing of the proximity induced gap in the normal metal due to 
the diffusion of normal electrons from the external circuit (which
plays the role of the N' reservoir) through the NN' boundary. For
nontransparent NN' boundaries ($\gamma _B=\infty $) this process
can be neglected and a real gap develops in the N-metal (Fig. 
\ref{lodo} (b)). As no such gap was found in \cite{gueron} we believe that
diffusion of normal excitations into the N-layer from the external
circuit should play an important role in these experiments. In other words,
the experimental situation appears to be closer to that described by the 
boundary condition $\gamma _B=0$ with a soft pseudogap (Fig. \ref{lodo}
(a)) 
than to the case $\gamma _B=\infty $ (see e.g. Ref. \onlinecite{belzig}).
The dependence of this effect on the size of the N-layer is depicted in Fig.
\ref{sizeeff}.

Making use of the Usadel equation one can easily recover simple
analytic expressions for the density of states at a distance $x_0$ 
away from the NS-boundary. For a N-wire of the total length $d$ at
$\epsilon\ll\epsilon_d$ we obtain
\begin{equation}
{N_N(\epsilon,x_0)\over
N(0)}=\alpha+{\epsilon^2\over\epsilon_d^2}\beta
\end{equation}
where $\alpha$ and $\beta$ describe the size effect
\begin{eqnarray*}
\alpha&=&\sin\left({\pi\over2}{x_0\over d}\right)\\
\beta&=&{32\over\pi^4}\alpha\left(1-{x_0\over
d}-\cos\left({\pi\over2}{x_0\over d}\right)\right)^2.\\
\end{eqnarray*}

Thus for $x_0\ll d$ the normal DOS at zero energy and $x=x_0$ is proportional
to $1/d$. Neglecting the charging effects (which in principle can
also be important \cite{gueron}) for the differential 
conductance of the tunneling probe we find
\begin{equation}
R_T\left.{dI\over dV}\right|_{V=0}=\alpha+{2\pi^2\over3}{T^2\over\epsilon_d^2}\beta
\end{equation}
These our results demonstrate that the depairing effect of the N'-reservoir 
needs to be taking into account on equal footing with pairbreaking due to 
inelastic scattering \cite{belzig,gueron}.

Let us also point out that one can also extract information about the 
correlation DOS by making two kinds of measurements with the same sample.
Indeed, by measuring the conductance of the system (or a
part of it) with no tunnel barriers one obtains information about the
combination of $N_N$ and $N_S$ entering the expression for the system 
conductance $G$, whereas
performing the tunnel experiments \cite{gueron} one probes only the normal
DOS $N_N$. Then the correlation DOS can be easily recovered.

\section{Electric field and charge}

In this section we shall discuss only the case of perfectly transparent 
interfaces.

\begin{figure}
    \centering
    \subfigure[Distribution of the electrical field]{\psfig{figure=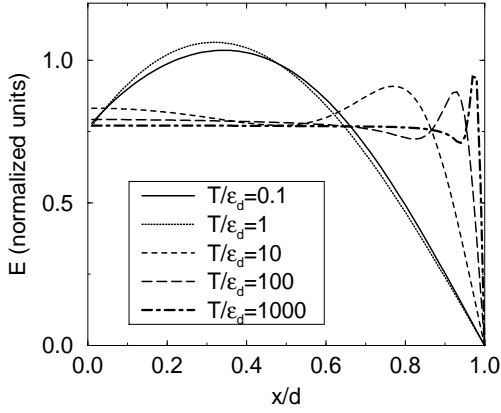,width=7cm,height=6cm}}

    \subfigure[Distribution of the electrical charge]{\psfig{figure=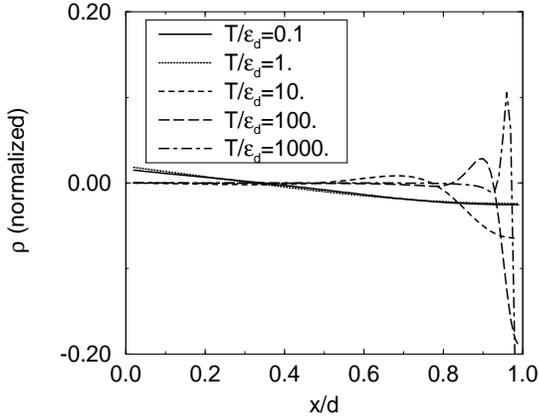,width=7cm,height=6cm}}
    \caption{Electrostatics within the wire}
    \label{rhofeld}
    \end{figure}

>From our solutions we can calculate the electric field and the charge by using
(\ref{potential}) and the Poisson's equation. The field shows essentially
non-monotoneous behavior. At $T=0$ we have
$E(\bar{x})=\cos(\bar{x}\pi/2)-{\pi\over2}(\bar{x}-1)\sin(\bar{x}\pi/2)$.
At high temperatures $T\gg\epsilon_d$, the field is constant $E=1$ far 
from the superconductor where no correlation remains
($1-\bar{x}\gg\xi_\epsilon$) and it changes linearly near the superconductor:
$E(\bar{x})=B_4(1-\bar{x})\sqrt{T\over\epsilon_d}$ with $B_4\approx2.59$, however, it still
overshoots in between these regimes (see Fig. \ref{rhofeld}). 

We see that close to the superconductor the electric field monotoneously
decreases with temperature as superconductivity becomes stronger there. 
Further from the NS boundary the field shows a complicated behavior 
overshooting the normal state value (the total voltage drop is fixed!)
in the region where superconducting correlation starts decaying either
due to thermal effects (high $T$) or due to the presence of a normal
reservoir (low $T$). The local resistivity is maximally lowered there 
and the layer of polarization charges is formed (see 
Fig. \ref{rhofeld}). 
These results emphasize again the importance of nonequilibrium effects
for understanding the behavior of the system conductance.
    
\section{Extension to systems containing a loop}

Recently, the properties of proximity wires containing a loop have
attracted much experimental \cite{Cour2,Cour3,Char} and theoretical \cite{Zai,Stoof3}
interest. 

    \begin{figure}
    \centerline{\epsfxsize8cm \epsffile{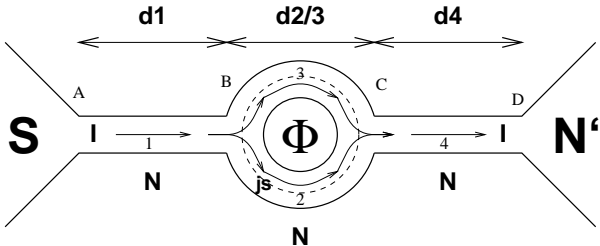}}
    \caption{The system under consideration. S and N are respectively
    superconducting and normal reservoirs. The wire is made of normal
    conducting material.}
    \label{loop}
    \end{figure}

If the wire was a real superconductor, the magnetic flux would induce
a supercurrent into the ring. As a function of $\Phi$, this current
has a period of the {\em superconducting\/} flux quantum $\Phi_0=h/2e$.

To describe these type of systems, our kinetic scheme has to be
extended in several points.

We define the Green's functions in the loop as
$$G=\cosh u_\epsilon\;\;\;\;F=\sinh u_\epsilon e^{i\varphi_\epsilon}\;\;\;
\varphi_\epsilon=\chi_\epsilon-2e\int_0^{\vec{x}} d\vec{l}\;\vec{A}(\vec{l})$$
where the integration goes along the loop. In the presence of a vector
potential, we have to introduce gauge independent derivatives
$$\nabla\longrightarrow\nabla-2ie\vec{A}.$$
This means, that instead of solving the Usadel equation with a vector
potential, we can perform a gauge transformation and map onto a system without
magnetic field having phase $\chi$ instead of $\phi$. As the
definition of the Green's functions has to be unique everywhere in the
loop, we have
$\lim_{x\rightarrow0+}\varphi(x)=\lim_{x\rightarrow0-}\varphi(x)\;(\mod
2\pi)$ or 
$$\lim_{x\rightarrow0+}\chi(x)-\lim_{x\rightarrow0-}\chi(x)={2e\Phi\over\hbar}\;(\mod
2\pi)$$ after gauge transformation. Here, $\Phi$ is the magnetic flux
in the ring. 

This mapping shows, that the magnetic field induces a supercurrent $j^S_\epsilon$
(screening current) into the system. We want to neglect any conversion
between this supercurrent and the dissipative current, so both are conserved seperately. This allows the application of
the kinetic scheme which
has been developed for systems without phase gradient \cite{VZK} but can be
generalized to any system where the dissipative current is conserved.

The Usadel equation then reads \cite{Z,WSZ}
\begin{equation}
{\cal D}{d^2\over dx^2}u_\epsilon=-2i\epsilon\sinh u_\epsilon+{{\cal D}\over2}\left(d\chi_\epsilon\over
dx\right)^2\sinh2u_\epsilon
\label{Usaphas}
\end{equation}

and has to be solved together with the equation for the conservation
of the supercurrent
\begin{equation}
{d\over dx}j_\epsilon^S=0\;\;\;\; j_\epsilon^S=|\sinh u_\epsilon|^2{d\chi_\epsilon\over
dx}
\label{Superc}
\end{equation}

In order to match the Green's functions at branching poits we use the 
standard continuity condition as no tunnel barriers are assumed to be there. 
>From the Usadel equation in matrix form 
$$ D\nabla(\check{g_\epsilon}\nabla\check{g_\epsilon})+i\epsilon[\tau_z,\check{g_\epsilon}]=0$$
follows for any branching point (see also \cite{Zai})
$$\sum_{i=1}^NA_i\check{g_\epsilon}{\partial\over\partial x_i}\check{g_\epsilon}=0$$
where the sum runs over matching branches, $\partial\over\partial
x_i$ denotes the derivative in the direction of branch $i$ and $A_i$
is the cross-section area of branch $i$. Using our
definitions, we get 
\begin{equation}
\sum_iA_i{\partial u_\epsilon\over\partial x_i}=0\;\;\;\;\sum_iA_i{\partial
\chi_\epsilon\over\partial x_i}=0
\label{branch}
\end{equation}
These conditions are equivalent to current conservation, so this is a
``Green's functions Kirchhoff law''. For $N=1$ Zaitsev's boundary
condition \cite{Zaibound}
for a Normal-Vacuum boundary is reproduced, $N=2$ is
equivalent to the trivial statement, that the Green's functions' 
derivatives are continuous within a branch.

For the calculation of the total transparency $D=1/m$, we can use the fact,
that the $m_i$ fulfill Ohm's law just by their definition:

$$m={{d_1m_1+d_4m_4+({1\over d_2m_2}+{1\over d_3m_3})^{-1}}\over(d_1+(1/d_2+1/d_3)^{-1}+d_4)}$$

\section{Magnetoresistance oscillations}
 
The equations (\ref{Usaphas}) and (\ref{Superc}) together with boundary
conditions (\ref{BC}) and branching conditions (\ref{branch}) have been solved
numerically and also analitically in some limiting cases. For
numerics, the problem 
was mapped onto a simpler
boundary value problem without any fitting point. 
As the system of equations is unstable, we used the relaxation
method \cite{numrec} instead of shooting. 

For convenience, we have chosen $d_1=d_2=d_3=d_4$ and
$A_1=2A_2=2A_3=A_4$, which simplifies the conditions \ref{branch}. The
effect of geometry on the conductance oscillations will be 
discussed in the appendix. The Thouless energy of just one branch will
be labeled as $\epsilon_d={D\over d_i^2}$.

\subsection{T-dependent Amplitude of $h/2e$-Oscillations} 

For $T=0$, only quasiparticles with the energy $\epsilon=0$ contribute
to the conductance. From \ref{Usaphas}
and \ref{BC} we can conclude, that $u_{\epsilon=0}$ is a purely
imaginary function, so the total conductance of the system is equal to
its normal state value, being independent of $\Phi$. In other words,
there exist no conductance oscillations at $T=0$ (cf. \cite{moriond,Stoof3}). 

At nonzero temperatures the system conductance depends on the 
magnetic flux inside the loop with the period equal to the
flux quantum $\Phi_0$ (see fig. \ref{period}). 
With the aid of simple analytic arguments
(see Appendix) one can conclude that at low temperatures the
amplitude of the conductance oscillations increases as $T^2$ (see
fig. \ref{magampl}).

\begin{figure}[]
     \centerline{\psfig{figure=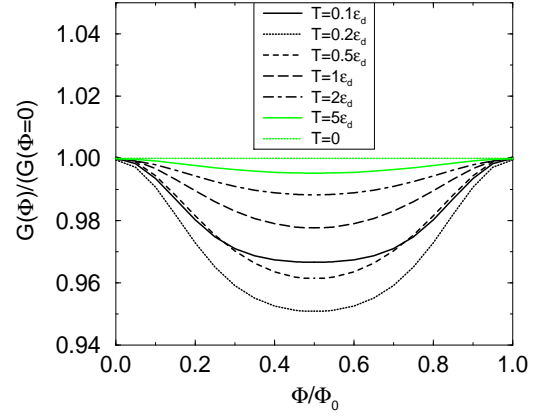,width=7cm,height=6cm}}
    \caption{$h/2e$-periodic structure of the conductance, normalized
    to the zero-field conductance}
    \label{period}
\end{figure}

In order to establish the temperature dependence of this amplitude
at higher $T \gg \epsilon_d$ it is convenient to make use of the fact
that for electrons with sufficiently large energies $\epsilon \gtrsim \epsilon_d$
superconducting correlation is destroyed already before they reach the 
loop. Thus at such energies the transparency of the whole structure $D(\epsilon)$
should be insensitive to the particular value of the flux inside the loop.
In other words, calculating the flux-dependent part of the system
conductance we can take into account only the contribution of low energy
quasiparticles which remain correlated in the loop area. E.g. for the
amplitude of the conductance oscillations we get
\begin{eqnarray}
\nonumber\Delta G(T)&=&G_{h/4e}(T)-G_{0}(T)\\
\nonumber&=&{1\over2T}\int_0^\infty d\epsilon\;
(D_{h/4e}(\epsilon)-D_{0}(\epsilon)) \sech^2(\epsilon/2T)\\
\nonumber&\approx& {1\over2T}\int_0^{\epsilon_{c}} d\epsilon\;
(D_{h/4e}(\epsilon)-D_{0}(\epsilon))
\sech^2(\epsilon/2T)\\
&\approx&{\epsilon_{c}\over2T}\Delta D_{av}
\label{oneovert}
\end{eqnarray}

where $\epsilon_{c}$ is the cutoff parameter of order $\epsilon_d$, and 
$\Delta D_{av}$ is constant. 

    \begin{figure}[]
     \centerline{\psfig{figure=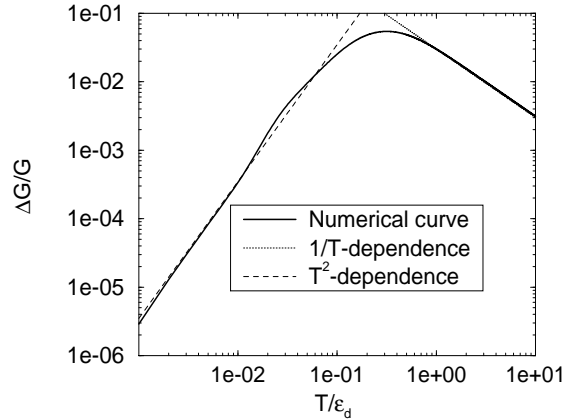,width=7cm,height=6cm}}
    \caption{Oscillation amplitude}
    \label{magampl}
    \end{figure}

\begin{figure}
\centering
\subfigure[Transparency]{\psfig{figure=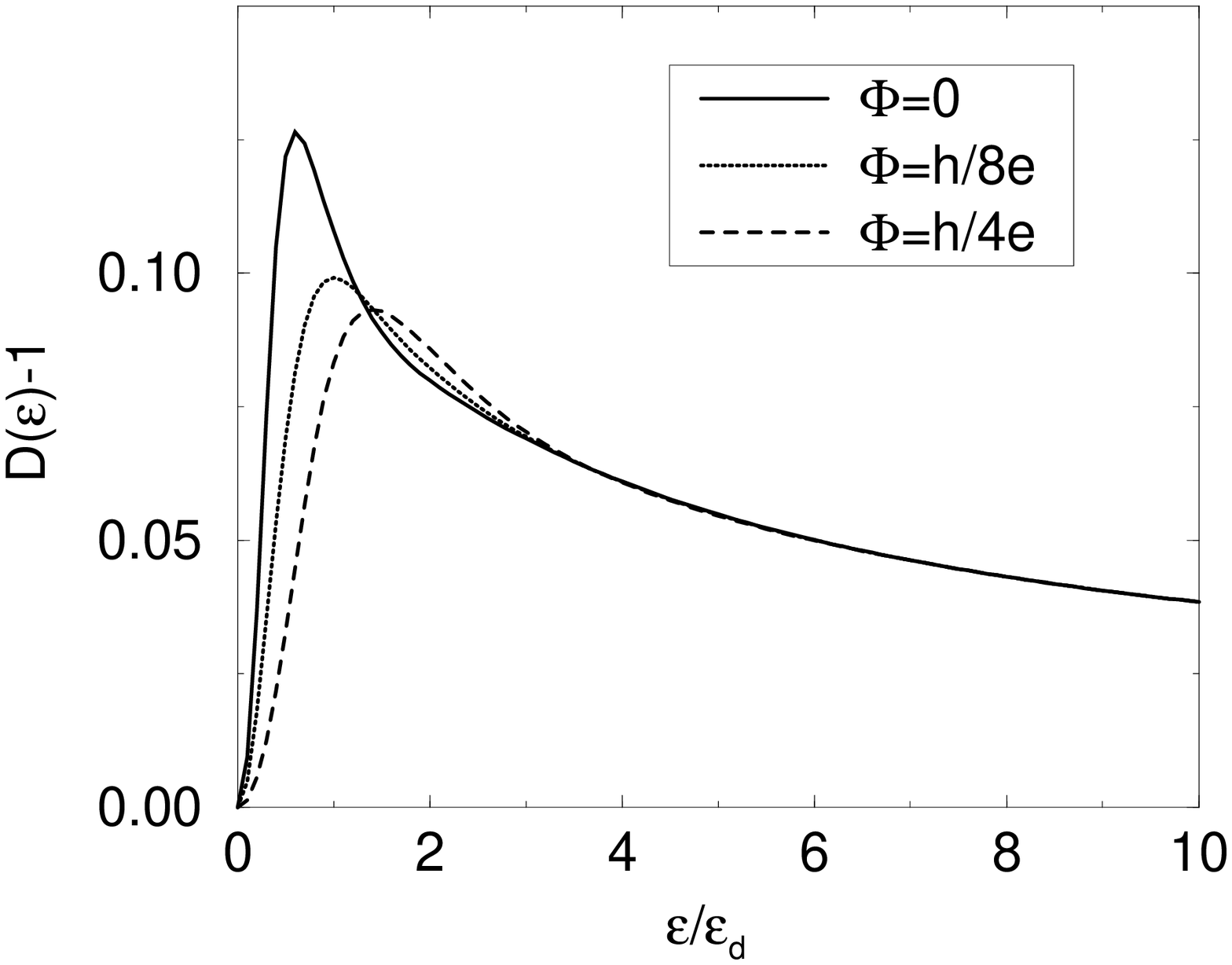,width=40mm,height=50mm}}\quad
\subfigure[Conductance]{\psfig{figure=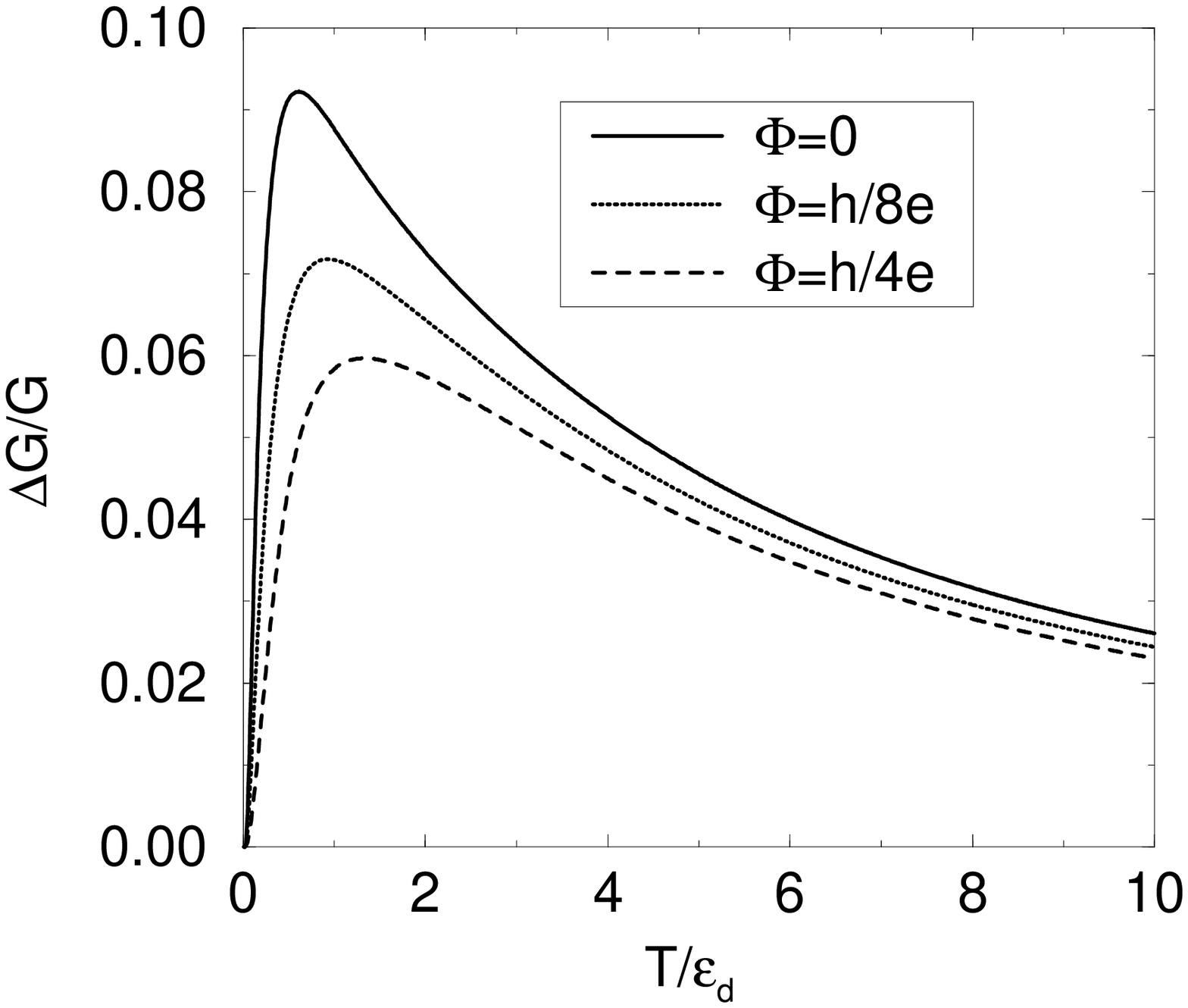,width=40mm,height=50mm}}
\caption{Temperature dependence of tranport properties at different fluxes}
\label{duenn}
\end{figure}

The results of our numerical analysis fully support the simple analytic
arguments presented above. The system transparency $D(\epsilon )$ 
is depicted in Fig. \ref{duenn} (a) for different values of the flux $\Phi$. 
The value $D(\epsilon )$ depends on $\Phi$ only at low energies, whereas
for $\epsilon \gtrsim \epsilon_d$ all curves merge. In accordance to 
our simple estimate (\ref{oneovert}) this leads to a $1/T$ decay of the 
oscillation amplitude $\Delta G$ at large $T$ (see Figs. \ref{duenn} (b) and
\ref{period}). Also the $T^2$ behavior of $\Delta G$ in the low temperature
limit is recovered (Fig. \ref{period}).

The $1/T$ behavior of $\Delta G$ has been also found in recent experiments
\cite{Cour2,Cour3}. We would like to point out that a slow power-law decay
of the conductance due to a dominating contribution of low energy
quasiparticles just emphasizes the physical difference between kinetic
and thermodynamic quantities, like supercurrent which decays exponentially
with increasing $T$. 

\subsection{Flux-Dependent DOS} 

As it was already discussed the simultaneous presence of correlated
electrons and the electric field in the normal metal causes
nontrivial nonequilibrium effects, the description of
which involves two densities of states $\nu (x)$ and $\eta (x)$.
In the presence of the normal metal loop with the magnetic flux $\Phi$
in our system there appeas a possibility to tune both normal and
correlation DOS by changing the value of $\Phi$. For the system 
depicted in Fig. \ref{loop} these densities of states can be easily
calculated. As one might expect for the region between the superconductor
and the loop (between the points A and B) this dependence is quite weak and
both DOS practically coincide with those calculated above for a wire without
the loop. On the other hand, in the region between the loop and
the normal reservoir N' (between the points C and D) the quantities 
$\nu (x)$ and $\eta (x)$ are very sensitive to the flux $\Phi$. 

\begin{figure}
\centering
\subfigure[Normal]{\psfig{figure=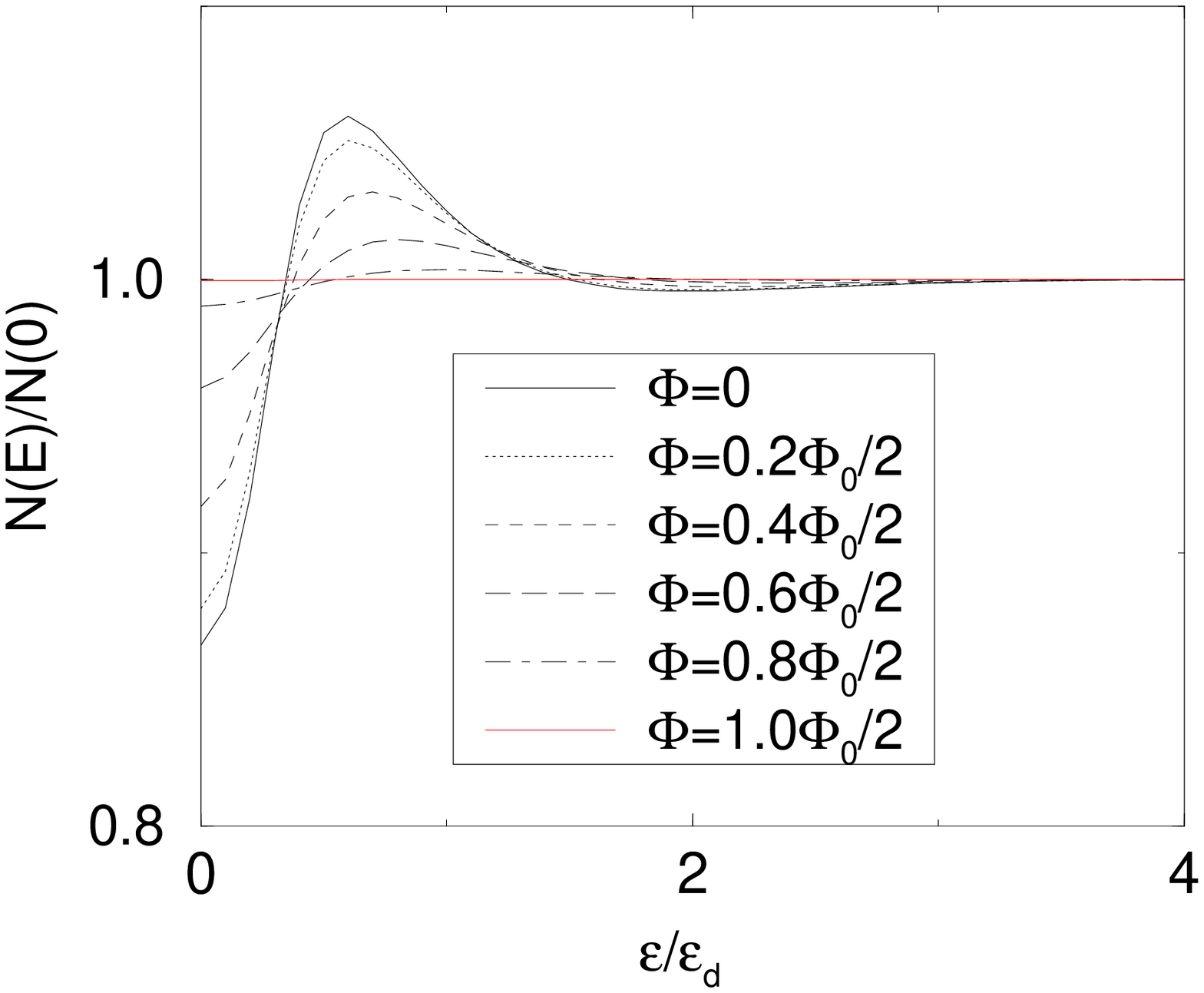,width=40mm,height=50mm}}\quad
\subfigure[Correlation]{\psfig{figure=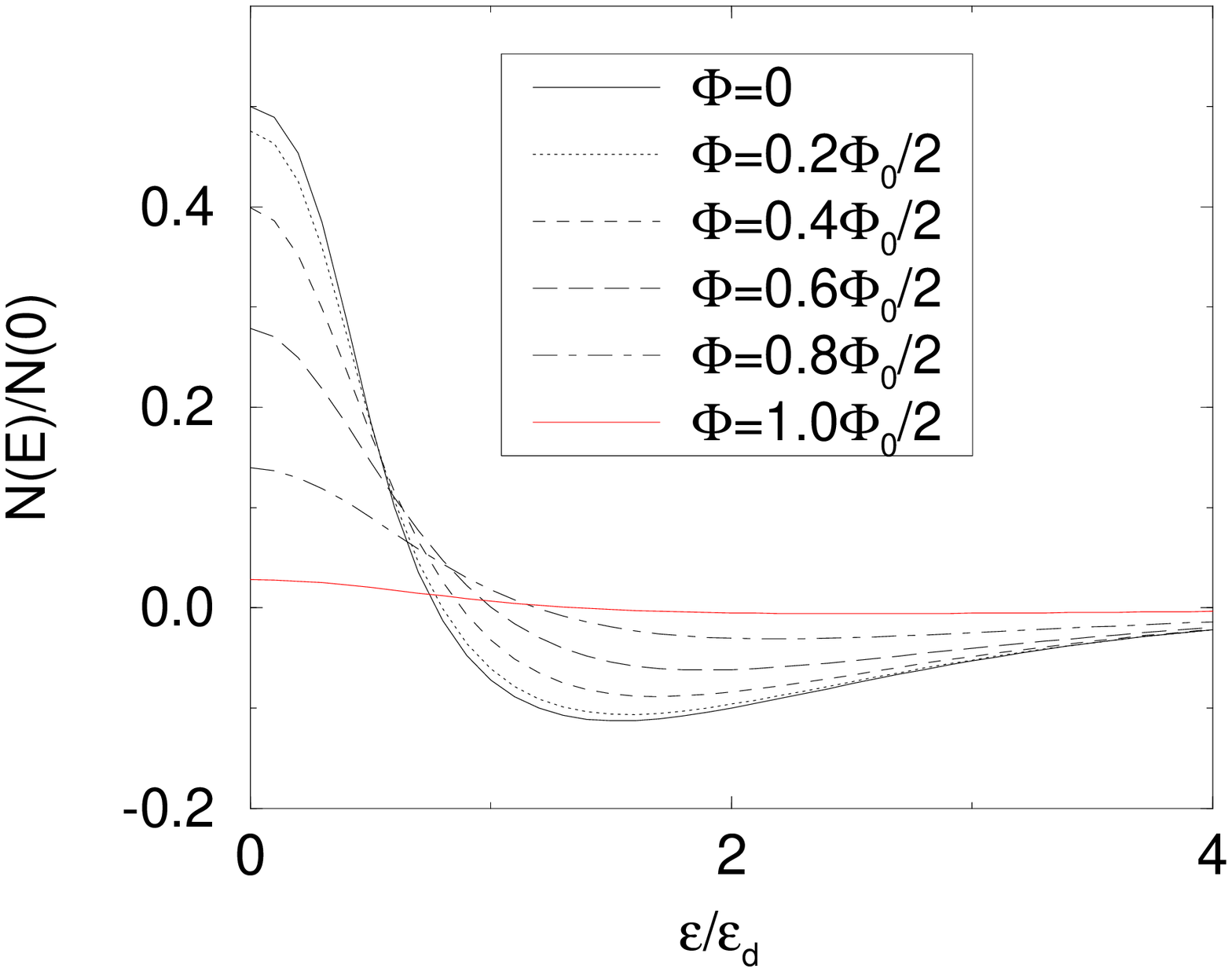,width=40mm,height=50mm}}
\caption{Flux-dependence of the two densities of states in point C of
the system}
\label{dosc}
\end{figure}

The normal and correlation DOS for the
point C (Fig. \ref{loop}) are presented in Fig. \ref{dosc}. We see that with 
increasing the value of the magnetic flux the proximity induced pseudogap
decreases and vanishes completely as the flux approaches the value $\Phi_0/2$.
For such value of $\Phi$ the proximity effect in the region ``after'' the
loop is completely destroyed, the pseudogap is fully suppressed and 
the normal DOS coincides with $N(0)$ at all energies. Accordingly the 
correlation DOS vanishes at $\Phi = \Phi_0/2$. Thus in this case
the resistance of the region between the points C and D is equal to its
normal state value at all $T$.

These results demonstrate that ``the strength'' of the proximity effect in our
system can be regulated by the external magnetic flux. This might serve as an
additional experimental tool for investigation of proximity induced
superconductivity in normal metalic structures. In particular we believe
that it would be interesting to repeat the tunneling experiments 
\cite{gueron} in the presence of the loop with the magnetic flux. Such 
experiments would provide a direct probe of the dependence of the 
densities of states on $\Phi$ (Fig. \ref{dosc} (a)).

\section{Summary and outlook}

We have used a microscopic kinetic analysis to describe the transport 
properties of superconductor-normal metal proximity structures. In 
the case of transparent inter-metallic boundaries we demonstrated 
a reentrant behavior of the system conductance with temperature. This 
behavior was attributed to nonequilibrium effects occuring in
the normal metal in the presence of proximity induced superconductivity
and the electric field. We argued that under these conditions 
both uncorrelated (``normal'') and correlated (``superconducting'') electrons
contribute to the system conductance which is henceforth defined by a 
combination of two densities of states -- the normal DOS $N_N$ and the 
correlation DOS $N_S$. The latter is known to play an important role
whenever the quasiparticle distribution function in a superconductor
is driven out of equilibrium \cite{Schmid}.

We studied the energy dependence of both these
quantities and demonstrated that if the normal metal is brought in a
direct contact to a superconductor on one side and a big normal reservoir
N' on the other side the normal DOS $N_N$ shows a soft pseudogap at energies
below the Thouless energy $\epsilon_d$. This effect is due to an interplay
between correlated and uncorrelated electrons penetrating into the N-layer 
respectively from a superconductor and a normal reservoir. If a
low transparency tunnel barrier is present at the NN' interface the
diffusion of normal excitations into the N-metal is suppressed, the
influence of a superconductor prevails and a real gap in the density
of states develops. 
 
Low transparent tunnel barriers also prevent the electric field from
penetration into the N-layer thus suppressing nonequilibrium effects
there. We demonstrated that with the aid of a proper combination
of the systems with and without tunnel barriers one can directly probe
both energy and spacial dependencies of both densities of states 
$N_N$ and $N_S$ in one experiment.

We extended our analysis to proximity systems containing the normal metal 
loop with the magnetic flux $\Phi$. We demonstrated that the conductance
of such systems as a function of $\Phi$ oscillates with the period equal
to the flux quantum $\Phi_0=h/2e$. The amplitude of these oscillations
$\Delta G$ also shows the reentrant behavior being equal to zero at $T=0$, increasing
as $T^2$ at $T \lesssim \epsilon_d$ and decaying as $1/T$ at 
$T \gtrsim \epsilon_d$. We argued that even at high temperatures $T \gg \epsilon_d$
low energy electrons with $\epsilon \lesssim \epsilon_d$ are only responsible for
the conductance oscillations leading to the power law decay of $\Delta G$ at
large $T$. We pointed out that the densities of states $N_N$ and $N_S$ can
be tuned (decreased) by applying the magnetic flux $\Phi$. In particular, if the flux in the loop is 
equal to the half of the flux quantum $\Phi =\Phi_0$ the proximity effect in
the region ``after'' the loop is completely suppressed, the normal DOS $N_N=N(0)$
at all energies and $N_S=0$. This effect can be also directly probed in tunneling
experiments and used for further studies of proximity induced superconductivity
in normal metallic systems.

We acknowledge useful discussions with C.Bruder, W.Belzig, H.Courtois, D.Esteve,
B.Pannetier, V.T.Petrashov, G.Sch\"on, B.Spivak, A.F.Volkov and B.J. van Wees.
This work was supported by the Deutsche Forschungsgemeinschaft
within the Sonderforschungsbereich 195. A.A.G acknowledges partial support by
RFFI No.96-02-1956.

\begin{appendix}

\section{A closer look at the $\hbox{h/2e}$-Oscillations}
Here we present further details related to the effect of geometry on
the behavior of the proximity NS systems containing a mesoscopic 
normal metal loop with the magnetic flux (see Fig. \ref{loop}).
In the first three sections
we will keep $A_1=2A_2=2A_3=A_4$ for simplicity allowing different
values for the $d_i$ but restricting ourself to symmetric loops $d_2=d_3$.

\subsection{Low temperature behavior}

For examining the low energy range, which is dominant for the
conductance oscillations at any temperature as stated above, we proceed perturbatively from the case
$\epsilon=0$, $\phi=0$ first to finite flux, then to finite
$\epsilon$. As the Usadel equation is
quadratic in $d\chi\over dx$, the value $u$ does not distinguish between the
upper and the lower branches of the ring (2 or 3 in Fig. \ref{loop}), so we will not make a difference in
the notation. 

We start from $\epsilon=0,\Phi=0$, where $u$ is purely imaginary and therefore
yields $D=1$. For a finite value of the flux but $\epsilon=0$, we get a purely imaginary correction and
therefore $D_{\phi,0}=1$. This correction is quadratic in the flux as
the r.h.s. of the equation is quadratic in the phase. The finite energy
correction at zero flux is a real function and is quadratic in $\epsilon$. 

Thus proceeding perturbatively we find $\Re
u \propto\epsilon\Phi^2$, and from the
expansion

\begin{eqnarray*}
u&=&-i{\pi x\over 2d_\Sigma}+i\phi^2g_{d_1,d_2,d_3,d_4}(x)+\epsilon
h_{d_1,d_2,d_3,d_4}(x)\\
&&+\epsilon\phi^2 k_{d_1,d_2,d_3,d_4}(x)+i\epsilon\phi^2
l_{d_1,d_2,d_3,d_4}(x)\\
\end{eqnarray*}

with $d_\Sigma=d_1+d_2+d_4$, in the leading order we get

\begin{eqnarray*}
D(\epsilon)&=&\left({1\over d_\Sigma}\int_0^{d_\Sigma} {dx\over\cosh^2(u_1(x))}\right)^{-1}\\
&=&\left({1\over{d_\Sigma}}\int_0^{d_\Sigma}
dx\left(1-{\epsilon^2\over2}(h+\Phi^2k)^2\right)+O(\epsilon^4,\phi^4)\right)^{-1}\\
&=&{1\over{d_\Sigma}}\left(1+{\epsilon^2\over2}(\lambda_{d_1,\dots,d_4}-\phi^2\mu_{d_1,\dots,d_4}\right)+O(\epsilon^4,\phi^4)\\
\end{eqnarray*}

where the coefficients are defined as

\begin{eqnarray*}
\lambda&=&\int_0^{d_\Sigma}dx h^2(x)\\
\mu&=&4\int_0^{d_\Sigma}dx h(x)k(x).\\
\end{eqnarray*}

Thus at low $T$ both the transparency and the conductance 
depend quadratically on energy and flux. Further analytic expressions
are presented in \cite{diplom}.

It is remarkable, that for $d_3=0$  due to (\ref{BC}) we have 
$\sinh u=0$ at the point $C$. Therefore, the current conservation condition 
(\ref{Superc}) can
be fulfilled for $j=0$, $\chi=\pm\phi/2$ (different signs refer to
different branches), so the phase gradient is zero almost everywhere
and the Usadel equation does not contain the phase any
more. Thus no magnetoresistance oscillations occur in this case.

\subsection{Cutoff energy}

Let us estimate the cutoff energy $\epsilon_{c}$.

\begin{figure}
\centering
\subfigure[Transparency]{\psfig{figure=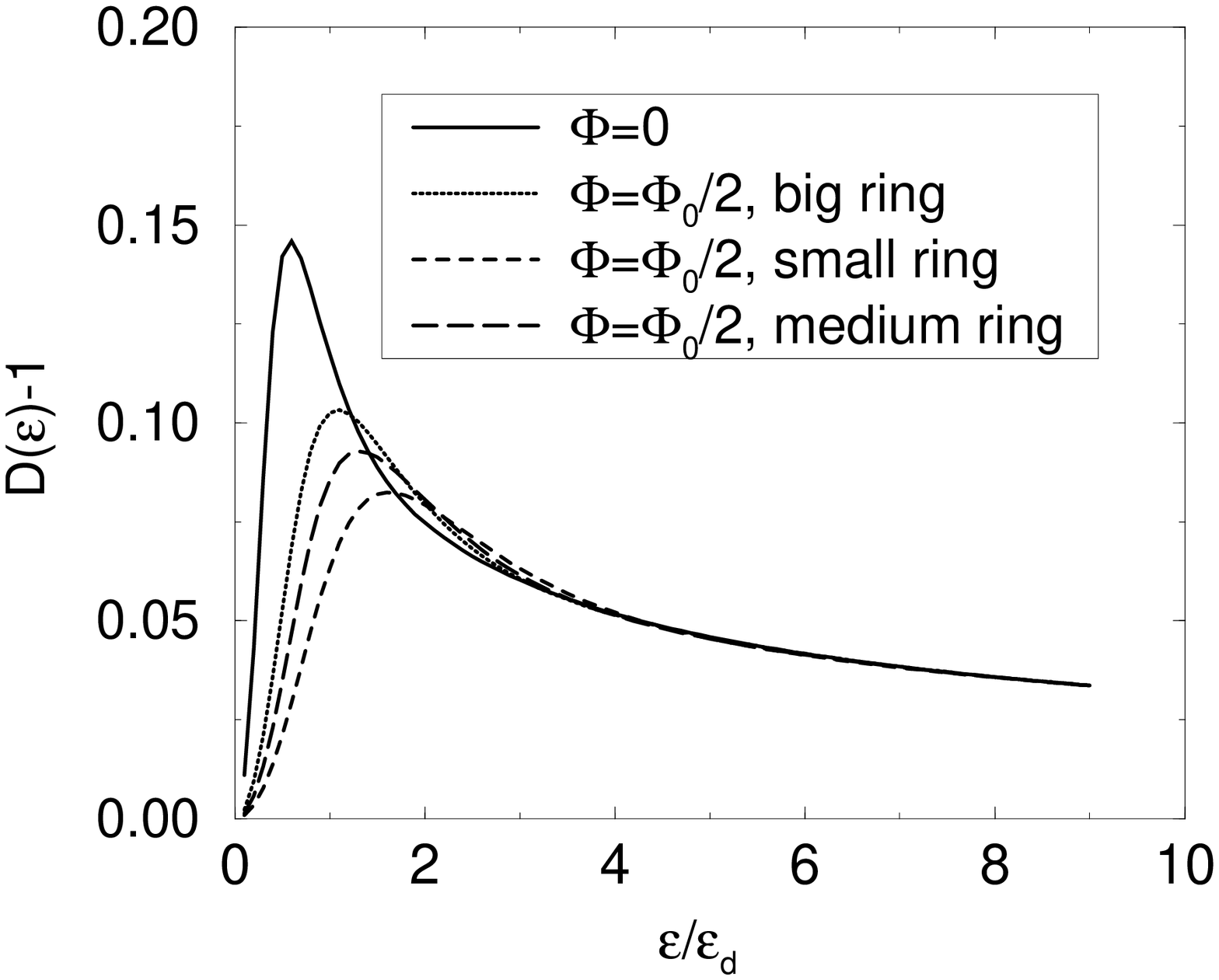,width=40mm,height=50mm}}\vspace{5mm}
\subfigure[Amplitude of conductance oscillations]{\psfig{figure=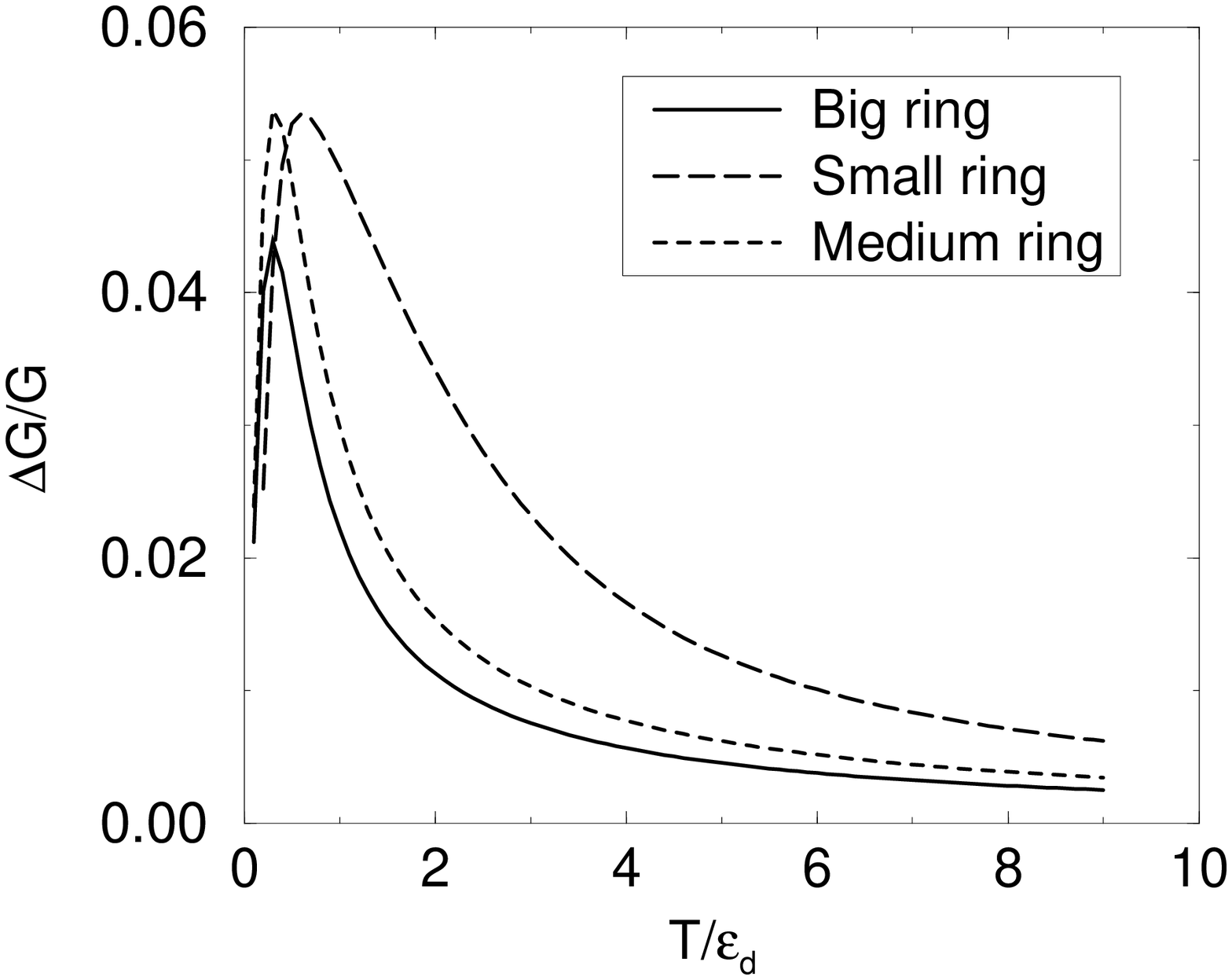,width=40mm,height=50mm}}
\caption{Size effects onto the conductance oscillations.
Small: $d_1=d_4=1.2$, $d_2=d_3=0.6$, Medium:$d_1=d_2=d_3=d_4=1$, Big:
$d_1=d_4=0.8$, $d_2=d_31.4$.}
\label{grkl}
\end{figure}

Consider the case $\epsilon\gg\epsilon_{d_1+d_2}={D\over
(d_1+d_2)^2}$. For $\Phi=0$ we again have (\ref{HochT})
$u=4\arctan \left({i\pi\over8}e^{-k^Rx}\right)$,
so for small $\phi$ we can proceed perturbutavely. As the supercurrent 
is exponentially small, we can approximate the phase profile as 
$${d\chi\over dx}={1\over2\sinh k_1d_2}e^{k_1(x-d_1-d_2/2)},$$
 so the influence of the magnetic
flux is concentrated within a distance
$\max\lbrace\xi_{N,\epsilon},d_2\rbrace$ from the point $C$ 
(see Fig. \ref{loop}). However, as $u$ is
exponentially small there, the oscillations of the transparency are exponentially
surpressed, so we can estimate $\epsilon_{c}\le \epsilon_{d_1+d_2}$,
which depends only on the sum of these length, but not on $d_1$ alone 
(see also fig. \ref{grkl}).

The key feature of Fig. \ref{grkl} (b) is the strong flux-dependence 
for systems with small rings. This fact can be also recovered from
the Usadel equation: for $d_2\ll d$ one estimates
${d\chi\over dx}\approx {\Phi\over2\Phi_0d_2}$. As this enters quadratically
and in the end we have to integrate over the ring only once, the
contribution of the ring is roughly $\propto 1/d_2$.

\subsection{Cross section effect}

\begin{figure}
    \centering
    \subfigure[Transparency]{\psfig{figure=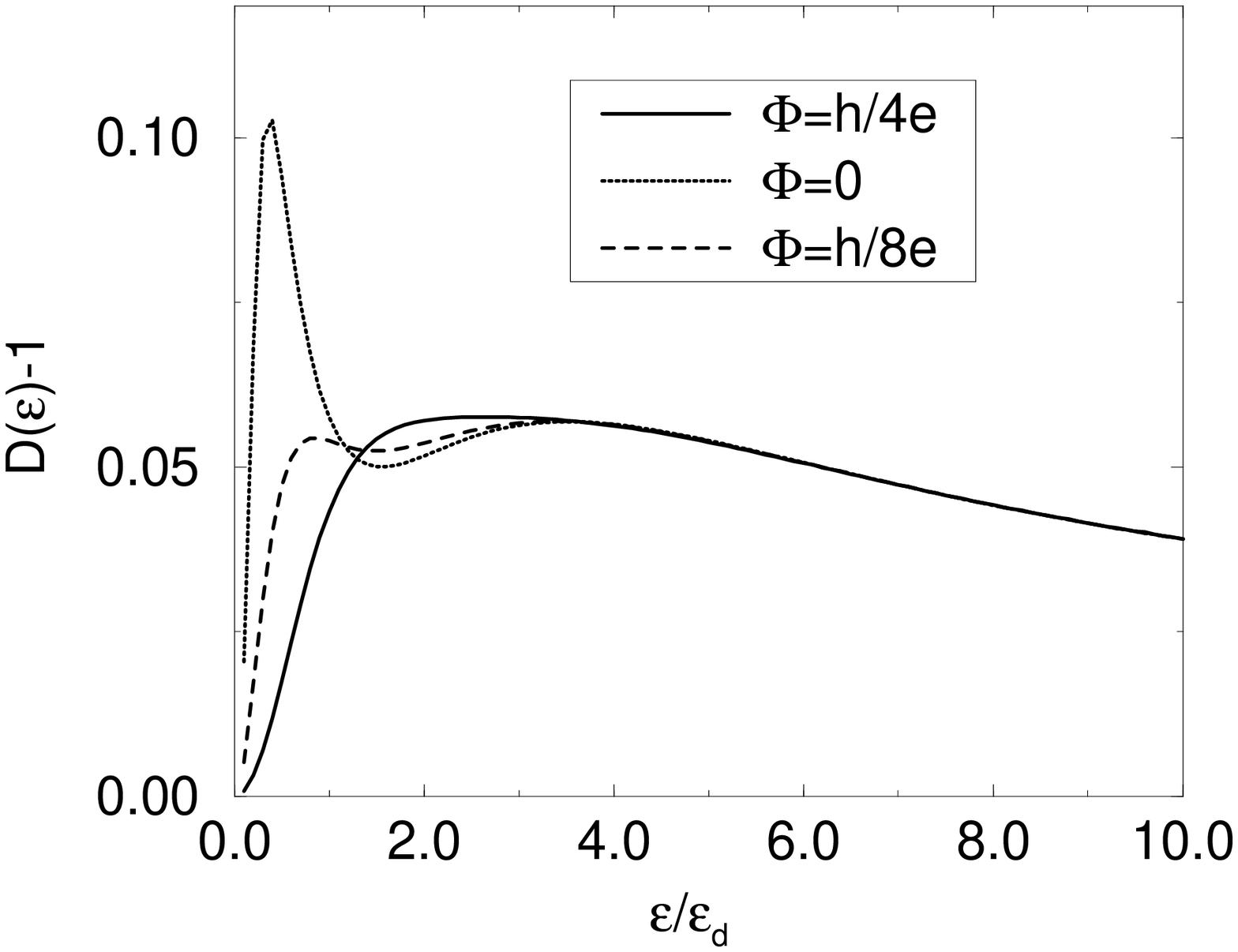,width=40mm,height=50mm}}\vspace{5mm}
    \subfigure[Conductance]{\psfig{figure=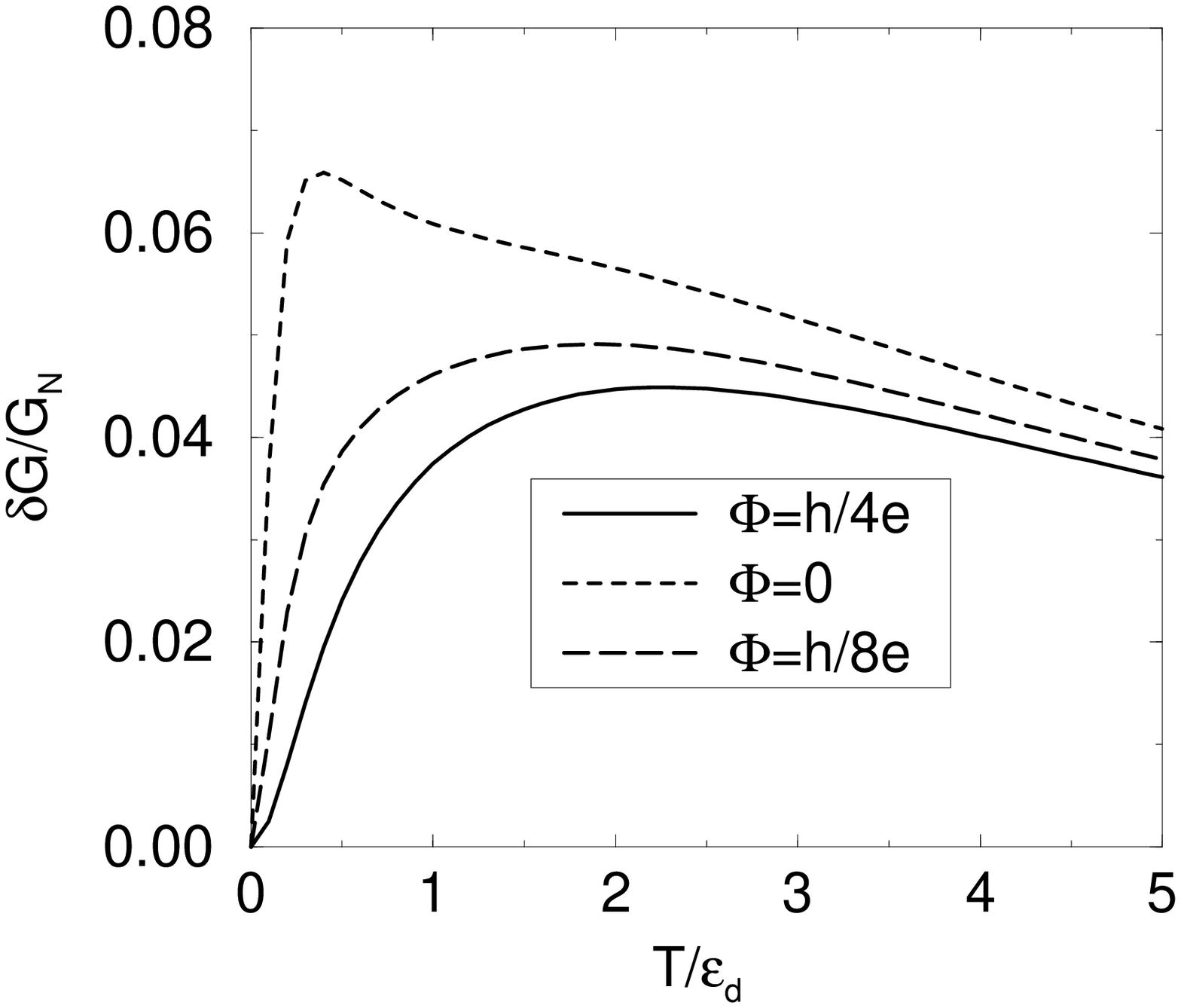,width=40mm,height=50mm}}
    \caption{Transport quantities for the system with $A_1=A_2=A_3=A_4$
    and $d_1=d_2=d_3=d_4$ displaying the cross-section effect.}
    \label{unter}
    \end{figure}

For the sake of simplicity above we have sticked to the case $2A_2=A_1=A_3$. 
As this condition might not be fulfilled in real experiments it is worthwhile to 
check whether the main features of our analysis survive for other values of 
$A_{1,2,3}$. In order to do that we performed calculations also for the case
$A_1=A_2=A_3$. The results are similar to those obtained
before, showing an additional
dip structure in the transparency at intermediate energies 
(Fig. \ref{unter} (a)) and a slightly deformed $G$ in the same energy interval
( Fig. \ref{unter} (b)). 

\end{appendix}


\begin{references}
\bibitem{FN} This argument still does not explain why at $T=0$ the system 
conductance is $exactly$ equal to its normal state value. 
Unlike in the case of ballistic NS constrictions (where a transparent physical
picture for the doubling of the linear conductance due to Andreev reflections
was presented \cite{BTK}) for diffusive NS structures a simple physical
interpretation of the above result -- if it exists at all -- is still lacking.
\end{references}
\end{document}